\documentclass[10pt,journal,cspaper,compsoc]{IEEEtran}

\usepackage{graphicx}
\usepackage{multibib}
\usepackage{multirow}
\usepackage{epstopdf}
\usepackage{amsmath,amssymb,,amsbsy}
\usepackage{subcaption,url,float,cite}
\usepackage[T1]{fontenc}
\usepackage[utf8]{inputenc}
\usepackage{authblk}
\usepackage[ruled,lined,linesnumbered,noend]{algorithm2e}
\usepackage{comment}
\usepackage[font=small,skip=0pt]{caption}
\usepackage{color}
\let\labelindent\relax
\usepackage{enumitem}

\newcommand{\mf}{\textsc{minTCO}}
\newcommand{\ms}{\textsc{minTCO-Perf}}
\newcommand{\mt}{\textsc{minTCO-RAID}}

\newcommand{\mff}{\textsc{minTCO-Offline}}

\makeatletter
\renewcommand\AB@affilsepx{, \protect\Affilfont}
\makeatother

\newcommand{\Brief}[1] {}

\begin{document}

\date{}

\title{\huge {I/O Workload Management for All-Flash Datacenter Storage Systems Based on Total Cost of Ownership{\thanks{This work was completed during Zhengyu Yang’s internship at Samsung Semiconductor Inc. This work was partially supported by National Science Foundation Career Award CNS-1452751 and AFOSR grant FA9550-14-1-0160.}}}}

\author[1]{\fontsize{10pt}{12pt}\selectfont Zhengyu Yang}
\author[2]{\fontsize{10pt}{12pt}\selectfont Manu Awasthi}
\author[3]{\fontsize{10pt}{12pt}\selectfont Mrinmoy Ghosh}
\author[1]{\fontsize{10pt}{12pt}\selectfont Janki Bhimani}
\author[1]{\fontsize{10pt}{12pt}\selectfont Ningfang Mi}

\affil[1]{Northeastern University, USA}
\affil[2]{Ashoka University, India}
\affil[3]{Samsung Semiconductor Inc., USA}

\setlength{\affilsep}{0.3em}
\renewcommand\Authfont{\normalfont}
\renewcommand\Affilfont{\itshape\small}
\renewcommand\Authands{ and }

\IEEEcompsoctitleabstractindextext{
\begin{abstract}
Recently, the capital expenditure of flash-based Solid State Driver (SSDs) keeps declining and the storage capacity of SSDs keeps increasing. 
As a result,  
all-flash storage systems have started to become more economically viable for large shared storage installations in datacenters, where metrics like Total Cost of Ownership (TCO) are of paramount importance. 
On the other hand, flash devices suffer from write amplification, which, if unaccounted, can substantially increase the TCO of a storage system.
In this paper, we first develop a TCO model for datacenter all-flash storage systems, and then plug a Write Amplification model (WAF) of NVMe SSDs we build based on empirical data into this TCO model. 
Our new WAF model accounts for workload characteristics like write rate and percentage of sequential writes. 
Furthermore, using both the TCO and WAF models as the optimization criterion, we design new flash resource
management schemes ({\mf}) to guide datacenter managers to make workload allocation decisions 
under the consideration of TCO for SSDs. 
Based on that, we also develop {\mt}  to support RAID SSDs and  {\mff} to optimize the offline workload-disk deployment problem during the initialization phase.
Experimental results show that {\mf} can reduce the TCO and keep relatively high throughput and space utilization of the entire datacenter storage resources. 
\end{abstract}

}

\maketitle

\IEEEdisplaynotcompsoctitleabstractindextext

\IEEEpeerreviewmaketitle

\vspace{-3mm}
\section{Introduction}
\label{SEC:IN}

The world has entered the era of ``Big Data'', when large amount of data is 
being collected from a variety of sources, including computing devices of all
types, shapes and forms. This data is then being pushed back to large, back-end 
datacenters where it is processed to extract relevant information.
As a result of this transformation, a large number of server-side 
applications are becoming increasingly I/O intensive. Furthermore, with the 
amount of data being gathered increasing with every passing day, the pressure on the I/O subsystem will continue to keep on increasing~\cite{wang2014bigdatabench}.

To handle this high I/O traffic, datacenter servers are being 
equipped with the best possible hardware available encompassing compute, memory,
networking and storage domains. Traditionally, I/O has been handled by hard 
disk drives (HDDs). HDDs have the benefit of providing an excellent economic value 
(\$/GB), but being built with mechanical moving
parts, they suffer from inherent physical throughput limitations, especially for random I/Os. To counter these performance limitations, solid state devices (SSDs) 
have recently begun to emerge as a viable storage alternative to HDDs.
In the recent past, SSDs have gained widespread adoption owing to reduced costs from the economies of scale. 
Datacenters, especially popular public cloud providers (e.g.,~\cite{amazon,google}) have been at the forefront of adopting flash technology.

Nevertheless, during this revolutionary change in cloud storage systems, 
flash-based SSDs face two major concerns: cost and write amplification (WA).
Firstly, the costs of owning (purchasing and maintaining) SSDs 
can still be very high. Balancing the trade-off between performance and economy is still an uphill battle.
Currently, Total Cost of Ownership (TCO),  comprising of two major costs, i.e., Capital and Operating Expenditures, remains a popular metric.
However, only a few prior studies have focused on the TCO of SSDs in datacenters, especially with the consideration of the cost of SSD's wornout. 

Secondly, SSDs have limited write cycles and also suffer from write amplification which is caused by a number of factors specific to flash devices including erase-before-rewrite, background garbage collection, and wear leveling. 
In fact, the Write Amplification Factor (WAF, will be defined in Sec.~\ref{SEC:WAF}) of an SSD is a direct function of the I/O traffic it experiences.
The I/O traffic, in turn, comprises of a number of different factors like the fraction of writes (as opposed to reads), the average size of I/O requests, the arrival rate of I/Os, and the ratio of sequential
I/O patterns (as opposed to random I/O) in the overall I/O stream. 
Greater WAF can significantly reduce the lifetime and increase the ownership cost of flash devices. 

Therefore, in this paper, we focus on the problem of deploying and allocating applications to a shared all-flash storage system of modern datacenters in order to reduce WAF and TCO. In detail, workloads (streams from an application) have different features, but in a long term of view, the I/O pattern of the same application can be characterized. Thus, we can address the above two concerns by investigating the relationship between workload patterns and WAF and then leveraging the relationship to develop new TCO models. 
In our experiments, we found that 
workloads with different sequential ratios have varying write amplifications even on the same SSD, which changes the lifetime of the device and eventually affects the TCO~\cite{mintcopp,mintcopt}. 
We are thus motivated to evaluate storage systems from a cost perspective that includes many dimensions such as maintenance and purchase cost, device wornout, workload characteristics, and total data amount that can be written to the disk, etc. 
Therefore, in this paper, we make the following contributions to achieve this goal.
\begin{itemize}[leftmargin=*]
\vspace{-1mm}
\item	We conduct real experiments to measure and characterize the write amplification under different
workloads, and reveal the relationship between write amplification and workload sequential ratio for each disk with fixed Flash Translation Layer (FTL) specs.
\vspace{-.5mm}
\item 	We propose a new TCO model while considering multiple factors like SSD lifetime, workload sequentiality, write wornout and the total writes.
\vspace{-.5mm}
\item 	We propose statistical approaches for calculating components that are essential for computing the TCO but cannot (practically) be measured from SSDs during runtime, such as write amplification and wornout of each SSD.
\vspace{-.5mm}
\item Based on our TCO model, we develop a set of new online adaptive flash allocation managers called ``{\mf}'', which leverage our TCO model to dynamically assign workloads to the SSD disk pool. The goals of {\mf} are: (1) to minimize the TCO, (2) to maximize client throughput as many as possible, and (3) to balance the load among SSD devices and best utilize SSD resources.
\vspace{-.5mm}
\item We present {\mt} to support RAID mode SSD arrays through an approximation approach. 
\vspace{-.5mm}
\item We develop {\mff} to support {\em offline} allocation scenario, where the datacenter manager needs to decide how many disks the datacenter needs and how to allocate workloads to the datacenter.
\end{itemize}

Lastly, we evaluate our new models and approaches using real world trace-driven simulation. Our experimental results show that {\mf} can reduce the TCO by up to 90.47\% compared to other traditional algorithms. 
Meanwhile, it guarantees relatively high throughput and spatial utilization of the entire SSD-based datacenter.
Moreover, {\mff} can also reduce TCO by up to 83.53\% TCO compared to the naive greedy allocation.

The remainder of this paper is organized as follows.
Sec.~\ref{SEC:WAF} investigates the cause of write amplification on SSDs.
Sec.~\ref{SEC:TCO} presents the details of our TCO models.  
Sec.~\ref{SEC:TCO_OA} proposes two versions of our {\mf} allocation algorithms. 
Sec.~\ref{SEC:EV}  measures the WAF of NVMe disks under different workload patterns, and evaluates our allocation algorithms. 
Sec.~\ref{SEC:RW} describes the related work.
Finally, we summarize the paper and discuss the limitations and future work plan of this research in Sec.~\ref{SEC:CO}.

\vspace{-3mm}
\section{Write Amplification Factor}
\label{SEC:WAF}
\begin{figure}[t]
	\centering
	\includegraphics[width=0.40\textwidth]{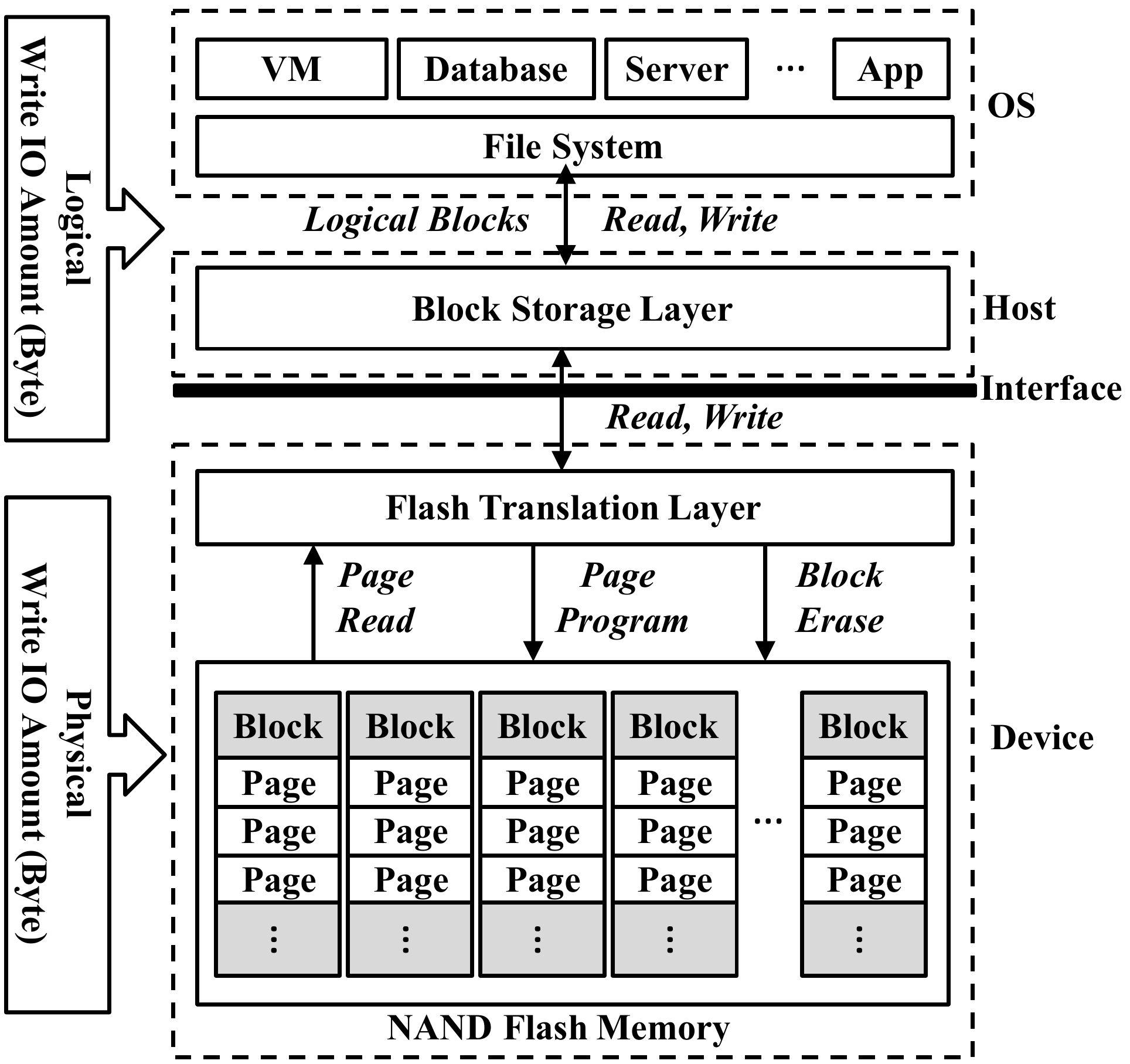}
	\caption{\small An example of I/O path from OS to device.}
	\vspace{-3mm} 
	\label{FIG:WAF_STRUCT}
\end{figure}

%
%
Write amplification factor (``$WAF$'', henceforth referred to as ``$A$'') is a commonly used metric to measure the write amplification degree. WAF is an undesirable phenomenon associated with flash devices where the actual amount of data written to the device is larger than the logical amount of data written by a workload. 
We define WAF as the ratio between the total physical write data written by the SSD and the total logical data written by the workload: $A=\frac{W_P}{W_L}$, 
%
%
where $W_L$ denotes the logical write amount (in bytes), and $W_P$ denotes the  
device-level physical I/O writes as seen by the SSD. Fig.~\ref{FIG:WAF_STRUCT} illustrates the logical and physical writes all the way from the application, through the OS, to the SSD. 
Large values of $A$ lead to increase I/O latency, shorten the SSD's
working lifetime, and increase power consumption. 

Where does the SSD write amplification come from? 
Flash devices have an unique property that they cannot be re-written unless they have been erased. Also, the minimum granularity of an erase operation is in the order of MBs (e.g., blocks), while the granularity of writes is much smaller, in the order of KBs (e.g., pages). Meanwhile, flash devices have limited write life cycles. Thus, for the purpose of wear-leveling, the logical address space in flash devices is dynamically mapped to the physical space and the mapping changes with every write. Flash devices have a software called FTL (Flash Translation Layer) running on them to manage the erase before re-write and wear-leveling requirements. The FTLs have to schedule periodic garbage collection events to de-fragment their write data. These garbage collection events can lead to extra writes that have not been generated by the host. Additionally, SSD reserves a user-invisible space (i.e., over-provision), which is helpful to reduce the WAF during these above-mentioned events to some extent. 
However, since flash devices have limited write-erase cycles, the mismatch between the two (logical and physical) types of writes can still cause the SSD to fail much more quickly than expected. 
On the other hand, besides these device hardware related factors, write amplification is also affected by I/O workload-related factors, such as mounted file systems and workload traffic patterns. 
Notice that in this paper,  we are not aiming to change SSD's FTL algorithms, instead, we mainly focus on the impact of different workload patterns to the WAF under SSDs (i.e.,  FTLs of SSDs are fixed in the datacenter after deployment).

\setlength{\textfloatsep}{10pt plus 1.0pt minus 2.0pt}
\vspace{-1mm}

The existing analytical models~\cite{peter} for WAF build the relationship between workload characteristics and WAF based on the different garbage collection policies (i.e., cleaning algorithms) and the impacts of the hot and cold data distribution~\cite{bux2010performance}. However, these models ignore a factor that is becoming increasingly important, especially in the NoSQL database community, traffic patterns of workloads in terms of sequential and random ratio experienced by the SSD~\cite{zhou2013characterizing}. 
With the proliferation of log structured merge tree (LSM tree) based NoSQL databases, there is a lot of uptick in the amount of sequential traffic being sent to the SSDs. 
LSM-tree based databases capture all the writes into a large in-memory buffer, and when the buffer is full, it is flushed to disk as a large, multi-gigabyte sequential write. 
Another similar case is write-intensive workloads that execute within virtual machines, where most of the write traffic to the SSD is usually sent out as large, sequential writes~\cite{tarasov2013virtual}. Hence, it is becoming increasingly essential to understand the WAF, performance of SSDs, device worn-out, and most importantly, the total owning cost of datacenters from a workload-centric view.

\vspace{-3mm}
\section{TCO Models of Storage Systems in All-Flash Datacenters}
\label{SEC:TCO}
	
\vspace{-1mm} 
TCO of an all-flash datacenter's storage system is a mix of a large number of items. Broadly speaking, these items can be broken down into two major categories: (1) Capital Expenditure (CapEx), and (2) Operating Expenditure (OpEx). Capital Expenditure refers to the amount of money that needs to be spent in setting up a facility. These include the cost of buying individual components of the servers, power supplies, racks that house the servers, among other things. OpEx, on the other hand, is the amount of money that is spent in the day-to-day operation of a datacenter. Examples of OpEx include electricity costs and personnel costs (required for maintaining the datacenter). CapEx is traditionally a large, one time expenditure~\cite{451researchcapexreport} while OpEx consists of small(er), recurring expenditures.
%
In this section, we develop a TCO model for an SSD intensive datacenter, based on the characteristics of the workloads (i.e., application I/O streams) that are scheduled on to those SSD devices. Our TCO model focuses specifically on the costs related to acquiring (i.e., CapEx) and maintaining (i.e., OpEx) SSDs in a datacenter.



\vspace{-3.8mm}
\subsection{Workload and Storage Models}
\label{SUBSEC:TCO_DSM}

\begin{figure}[t]
\centering
\includegraphics[width=0.32\textwidth]{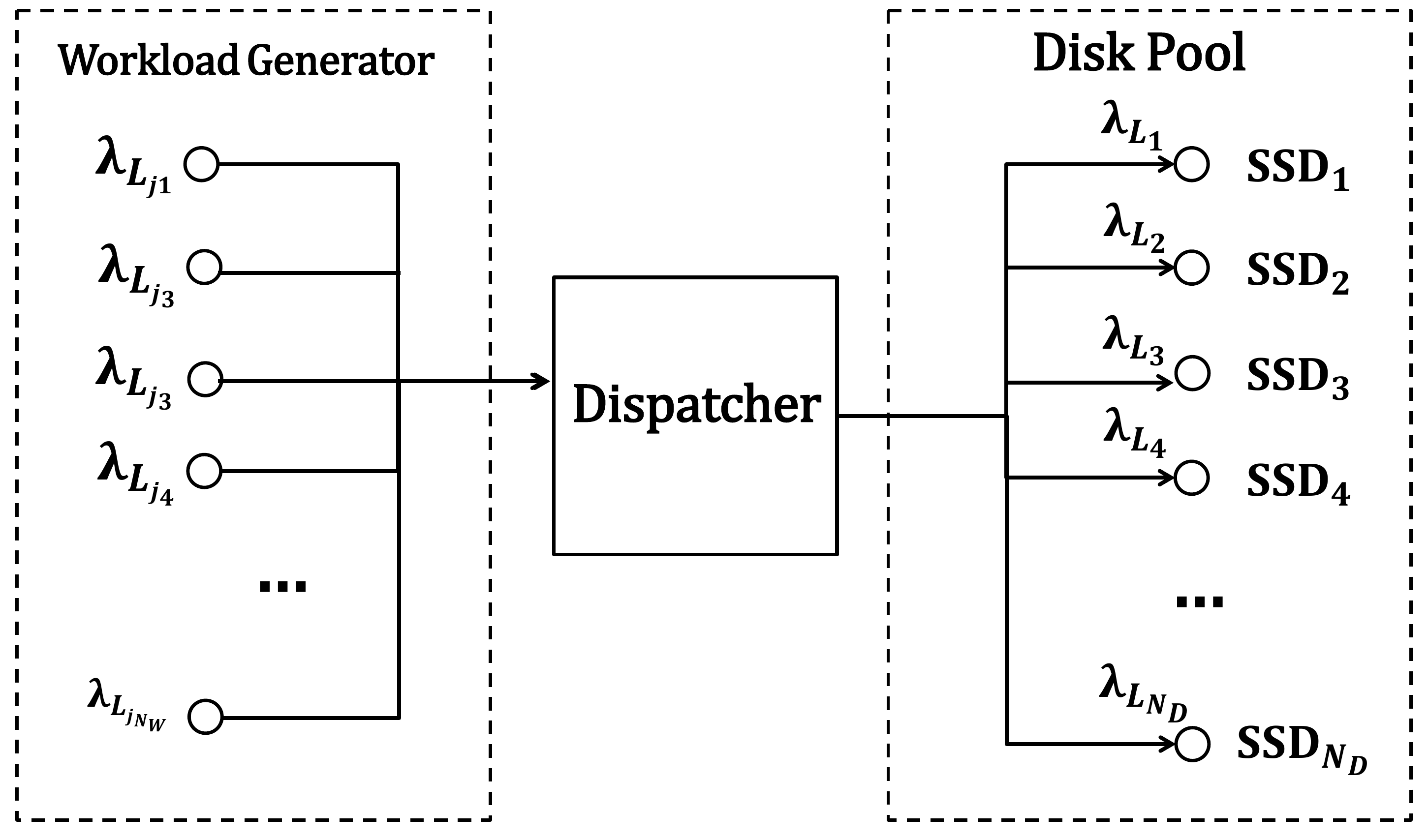}
\caption{\small Model of a datacenter storage system.}
\label{FIG:TCO_QUEUE}
\end{figure}

\vspace{-1mm} 
First, we briefly explain our assumptions about datacenters, their workloads and storage systems.
We  assume the datacenter storage system to be a large pool of SSD devices. This helps us abstract the problem of modeling SSDs from a per-server resource to a pool of datacenter-wide resources. 
We then model the storage system of such a datacenter as a {\em workload-to-disk allocation
problem}, as shown in Fig.~\ref{FIG:TCO_QUEUE}.
In this model, we have a pool of $N_D$ SSDs as shown in Fig.~\ref{FIG:TCO_QUEUE}.
Meanwhile, there are $N_W$ applications (workloads) that submit I/O requests with logical write rates $\lambda_{{L_J}_{i}}$ (where $1 \le i \le N_W$, and ``$L_J$'' stands for ``logical'' and ``job''), as seen in the left hand box of Fig.~\ref{FIG:TCO_QUEUE}. 
%
To complete the connection between I/O requests and the SSDs, we assume an allocation algorithm that is used by the dispatcher to assign I/O workloads to different SSDs. 
Each workload has its own characteristic, and arrives at the dispatcher at different times.
Multiple workloads can be assigned to the same disk as long as the capacity (e.g., space and throughput) of the disk is sufficient, 
such that the logical write rate ($\lambda_{L_i}$) to disk $i$ is the summation of logical write rates from the workloads in the set $\mathbb{J}_i$ that are allocated to that SSD, i.e., $\lambda _{ L_{ i } }=\sum _{ j\in \mathbb{J}_{ i } }^{  }{ \lambda _{ {L_J}_j } }$.
We summarize our main assumptions in the following subsections.
\vspace{-2mm} 

\subsubsection{\textbf{I/O Workload with Certain Properties}}
``Workload'' is defined as an endless logical I/O stream issued by applications. Particularly, from a long-term view (e.g., years), characteristics of workloads, such as sequential ratio, daily write rate, read-write ratio, working set size, and re-access ratio, can be abstracted as (almost) fixed values.  Workloads may arrive to the datacenter at different times. Once a workload arrives, the dispatcher assigns it to one certain disk (or multiple disks for RAID mode SSDs, see Sec.~\ref{SUBSEC:TCO_OA_BS}), and the disk(s) will execute this workload until the disk(s) is (are) ``dead'' (i.e., SSD write cycle limit is reached), or the workload finishes. We ignore the overhead (such as time and energy consumption) during the workload deployment. 

\setlength{\textfloatsep}{10pt plus 1.0pt minus 2.0pt}

\vspace{-2mm} 

\subsubsection{\textbf{Isolation among Multiple Workloads on a Single SSD}}
Multiple workloads can be assigned to a single SSD, and have separate and independent working sets (i.e., address spaces and segments are isolated). Therefore, the cross-workloads effects along I/O path due to interleaving working sets are negligible.

\vspace{-2mm} 

\subsubsection{\textbf{SSD's Write Amplification Model}}
We use the WAF model to capture the behavior of an SSD under a workload with a specific I/O pattern. Our WAF model can estimate the WAF of each disk by using the sequentiality information of multiple workloads that are concurrently executing at a particular SSD. The write wornout of each SSD can further be estimated using the runtime status of that SSD's WAF.



\vspace{-3mm}
\subsection{Total Cost of Ownership Model}
\label{SUBSEC:TCO_TM}

\begin{figure*}	
	\centering
	\includegraphics[width=0.85\textwidth]{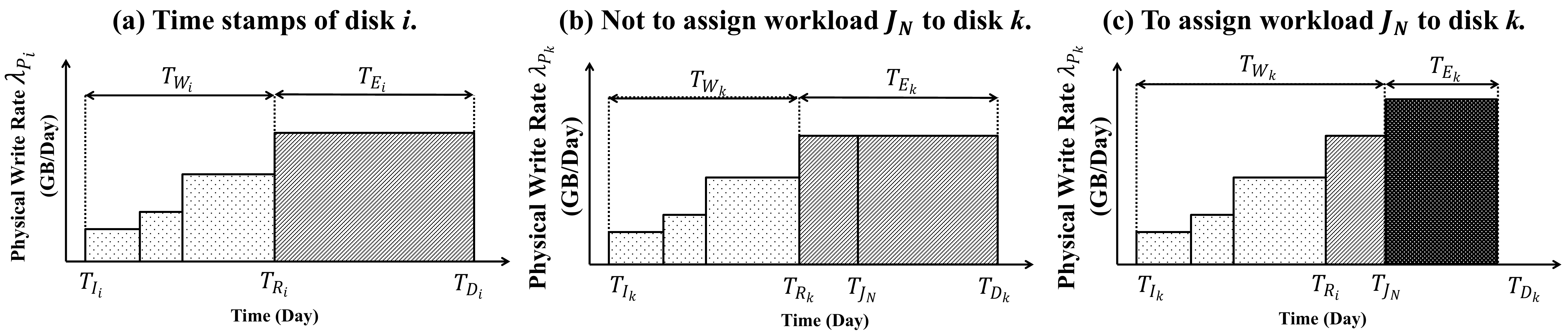}
	\caption{\small Examples of physical write rates vs. workload arrivals.} 
	\label{FIG:TCO_TIME}
\end{figure*}

Owning and maintaining a low cost SSD-intensive datacenter is critically important. TCO has been widely adopted to evaluate and assess storage subsystem solutions for traditional hard drives.
However, to the best of our knowledge, there is no standard formula for calculating the TCO of the SSD-intensive storage subsystem.
In order to comprehensively access the expenditure of a datacenter, a generic TCO model should consider purchasing and maintenance costs, service time, served I/O amount and device wornout.
We present the following models to calculate the TCO.
As we mentioned, two major types of costs: CapEx ($C_{I_i}$) and OpEx ($C_{M_i}$) are considered in the basic TCO model. In detail,  $C_{I_i}  =C_{Purchase_i}+C_{Setup_i}$ and ${C'}_{M_i} =C' _{Power_i}+C'_{Labor_i}$, 
where $C_{Purchase_i}$ and $C_{Setup_i}$ are one-time cost (\$) of device purchase and device setup, and $C' _{Power_i}$ and $C'_{Labor_i}$ are power and maintenance labor cost rate (\$/day). 
Although CapEx (${C'}_{I_i}$) is one time cost, OpEx ($C'_M$) is a daily rate and the TCO should be depend on the amount of time that an SSD has been used. Therefore, we need to attach a notion of \textit{time} to TCO. Assume we know the expected life time ($T_{{Lf}_i}$) of each disk (i.e., $T_{{Lf}_i}=T_{D_i}-T_{I_i}$, where $T_{D_i}$ and $T_{I_i}$ are the time when the disk $i$ is completely worn out and the time when it was started accepting its first request, respectively), the total cost for purchasing and maintaining a pool of SSDs can be calculated as:
\vspace{-1.5mm} 
\small
\begin{eqnarray}
TCO=\sum _{ i=1 }^{ N_D }  {  (     C_{I_i} + C' _{M_i}  \cdot   T_{Lf_{i}}   ) },\vspace{-2mm} 
 \label{EQ:TCOB}
\end{eqnarray}
\normalsize
where $N_D$ is the number of disks in the pool. Fig.~\ref{FIG:TCO_TIME}(a) also illustrates an example from time stamp view, where I/O workloads keep arriving and thus the physical write rate of disk $i$ increases accordingly. 
However, Eq.~\ref{EQ:TCOB} does not reflect SSD wornout at all, which is highly coupled with the workload arrival distribution.
For instance, in a datacenter consisted of the same type of SSDs, the SSD running workloads with the highest physical write rate may always have the highest TCO value (i.e., its $C' _{M_i}  \cdot   T_{Lf_{i}}$ is the greatest among all) according to Eq.~\ref{EQ:TCOB}, since this SSD is probably the last one to be worn out. However, this SSD may have a larger WAF due to the workload pattern, while others that died earlier may have smaller WAFs and can serve more logical write I/O amounts under the same cost.
Therefore, to conduct a fair and meaningful comparison, we introduce the data-averaged TCO rate ($TCO'$) from the perspective of the cost vs. the total amount of (logical) data served (written) to an SSD as follows.
\vspace{-2mm} 
\small
\begin{eqnarray}
TCO'=\frac { \sum _{ i=1 }^{ N_{ D } }{ (C_{ I_i }+C' _{ M_{ i } } \cdot T_{{Lf}_i}) }  }{ \sum _{ j=1 }^{ N_{ W } }{ D_j }  } ,\vspace{-1mm} 
\label{EQ:TCORD}
\end{eqnarray}
\normalsize
where $\sum _{ j=1 }^{ N_{ W } }{ D_j }$ is the total {\em logical} data write amount for all $N_W$ workloads.
Again, we use logical writes as a proxy for physical writes not only because the former is much easier to obtain for most workloads, but also because by being normalized by the logical writes, the $TCO'$ is able to reflect the WAF and judge the disk-level wear leveling performance of different allocation algorithms.

\setlength{\textfloatsep}{10pt plus 1.0pt minus 7.0pt}
\vspace{-3mm}
\subsection{Calibrating TCO Models}
\label{SUBSEC:TCO_CTM}

\vspace{-1mm}
The models developed in the prior section have all assumed that certain parameters 
for TCO calculation (e.g., total logical data write amount, expected lifetime, etc.) 
are readily available or measurable. However, it is impractical to measure some parameters that are necessary in our TCO models. 
In this section, we propose a mathematical approach of estimating those hard to be directly measured parameters. 

\vspace{-1mm}
\subsubsection{\textbf{Total Logical Data Writes $\sum _{ j=1 }^{ N_{ W } }{ D_j } $}}
\label{SUBSUBSEC:TCO_CTM_TLDW}

Given workload $j$'s logical write rate $\lambda_{L_j}$, arrival time $T_{A_j}$ and 
estimated time of death ($T_{D(j)}$) of workload $j$ 's host disk $D(j)$, we can calculate the total amount of
data written by all the jobs over their course of execution as: $\sum _{ j=1 }^{ N_{ W } }{ D_j } = \sum _{ j=1 }^{ N_{ W } }{ \lambda _{L_ j } (T_{D(j)}-T_{A_j}) }$,
where $\lambda_{L_j}$ is the logical data write rate of workload $j$.
The only unknown parameter left is $T_{D(j)}$, which can be obtained by calculating each host disk's expected lifetime.

\vspace{-1mm}
\subsubsection{\textbf{Expected Lifetime $T_{{Lf}_i}$}}
\label{SUBSUBSEC:TCO_CTM_EL}

The lifetime of a disk depends not only on the write traffic from the currently executing
jobs, but also on those jobs that have already been deployed on the disk. Furthermore, we also need to account
for the effects of the \textit{combined} write traffic of the workloads that are concurrently
executing on a disk.
As shown in Fig.~\ref{FIG:TCO_TIME}(a), the lifetime of disk $i$ is the period from $T_{I_i}$ to $T_{D_i}$.
We further split the lifetime into two phases: (1) all (accumulated) working epochs ($T_{W_i}$) until the last workload arrives, i.e., $T_{W_i}=T_{ R_i }-T_{I_i}$, where $T_{ R_i }$ is the assigned time of the most recent workload; and (2) expected work lifetime
($T_{E_i}$) from $T_{ R_i }$ to the expected death time, i.e., $T_{E_i}=T_{ D_i }-T_{R_i}$. 
The former is easy to monitor, and the latter is further estimated as the available remaining write cycles of disk $i$ divided by the physical data write rate ($\lambda_{P_i}$) of disk $i$ from $T_{Ri}$. Moreover, $\lambda_{P_i}$ can be calculated as disk $i$'s logical data write rate ($\lambda_{ L_i }$) times disk $i$'s WAF ($A_i $). 
Thus, we have $T_{{Lf}_i}=T_{D_i}-T_{I_i}=T_{W_i}+T_{E_i}
=(T_{ R_i }-T_{I_i})+\frac { W_i - w_i }   { \lambda_{ P_i }  }
=(T_{ R_i }-T_{I_i})+\frac { W_i - w_i }   { \lambda_{ L_i } \cdot A_i }
=(T_{ R_i }-T_{I_i})+\frac { W_i - w_i }   { \lambda_{ L_i } \cdot f_{seq}(S_i) }$,
where $A_i$, $W_i$, $w_i$ and $S_i$ are the WAF function, the total write limit, current write count (wornout), and sequential ratio of all running workloads of disk $i$, respectively. Since the SSDs' hardware are fixed, we denote $A_i$ as a function of workload's write I/O sequential ratio ($f_{seq}$) of disk $i$, which will be validated and regressed in our experimental section (Sec.~\ref{SUBSUBSEC:EV_WAMM_WRM}). In fact, we can plug any WAF model into this TCO model.
As of now, we also know $T_{R_i}$, $T_{I_i}$ and $W_i$, and what we need to estimate next are the remaining parameters, i.e., $\lambda_{L_i}$, $S_i$ and $w_i$.

\subsubsection{\textbf{Logical Write Rate of Workloads on Disk $\lambda_{L_i}$}}
\label{SUBSUBSEC:TCO_CTM_LWRW}

For disk $i$, its logical write rate $\lambda_{L_i}$ should be the sum of all its assigned workloads' logical write rates, i.e., ${ \lambda  }_{ { L }_{ i } }=\sum _{ j\in { \mathbb{J} }_{ i } }^{  }{ { \lambda  }_{ { L }_{ ij } } }$,
where $\mathbb{J}_i$ is the set of workloads running on disk $i$. Notice that there is a boundary case
during the very early stage when no workloads have been assigned to the disk $i$ (i.e., $\mathbb{J}_i=\emptyset$), such that 
$\frac { W_i - w_i }   { \lambda_{ L_i } \cdot f_{seq}(S_i) }$ becomes infinite. To avoid such an extreme case, we conduct a warming up process that assigns at least one workload to each disk. Only after this warming up phase is done, we start to calculate $T_{Lf_i}$.

\vspace{-1mm}
\subsubsection{\textbf{Sequential Ratio of Workloads on Disk $S_i$}}
\label{SUBSUBSEC:TCO_CTM_SRW}

In order to calculate the write amplification $A_i$ in Sec.~\ref{SUBSUBSEC:TCO_CTM_EL}, we need to know the sequential ratio of multiple workloads that are assigned to one disk. Unlike the logical write rate, the combined sequential ratio of multiple workloads is not equal to the sum of sequential ratios of all workloads. Our estimating solution is to assign a weight to each workload stream's sequential ratio and set the weight equal to the workload's logical data write rate. Hence, for multiple workloads running on the disk, we can calculate the overall sequential ratio as: ${ S }_{ { i } }=\frac { \sum _{ j\in { \mathbb{J} }_{ i } }^{  }{ { \lambda  }_{ { L }_{ ij } }S_{ { ij } } }  }{ \sum _{ j\in { \mathbb{J} }_{ i } }^{  }{ { \lambda  }_{ { L }_{ ij } } }  }$.
where $\lambda_{L_{ij}}$ and $S_{ij}$ are the logical write rate and  sequential ratio of $j$th workload running on disk $i$.
Appendix 1 shows the details of our implementation of sequential ratio estimator.

\begin{figure}[t]
	\centering
	\includegraphics[width=0.4\textwidth]{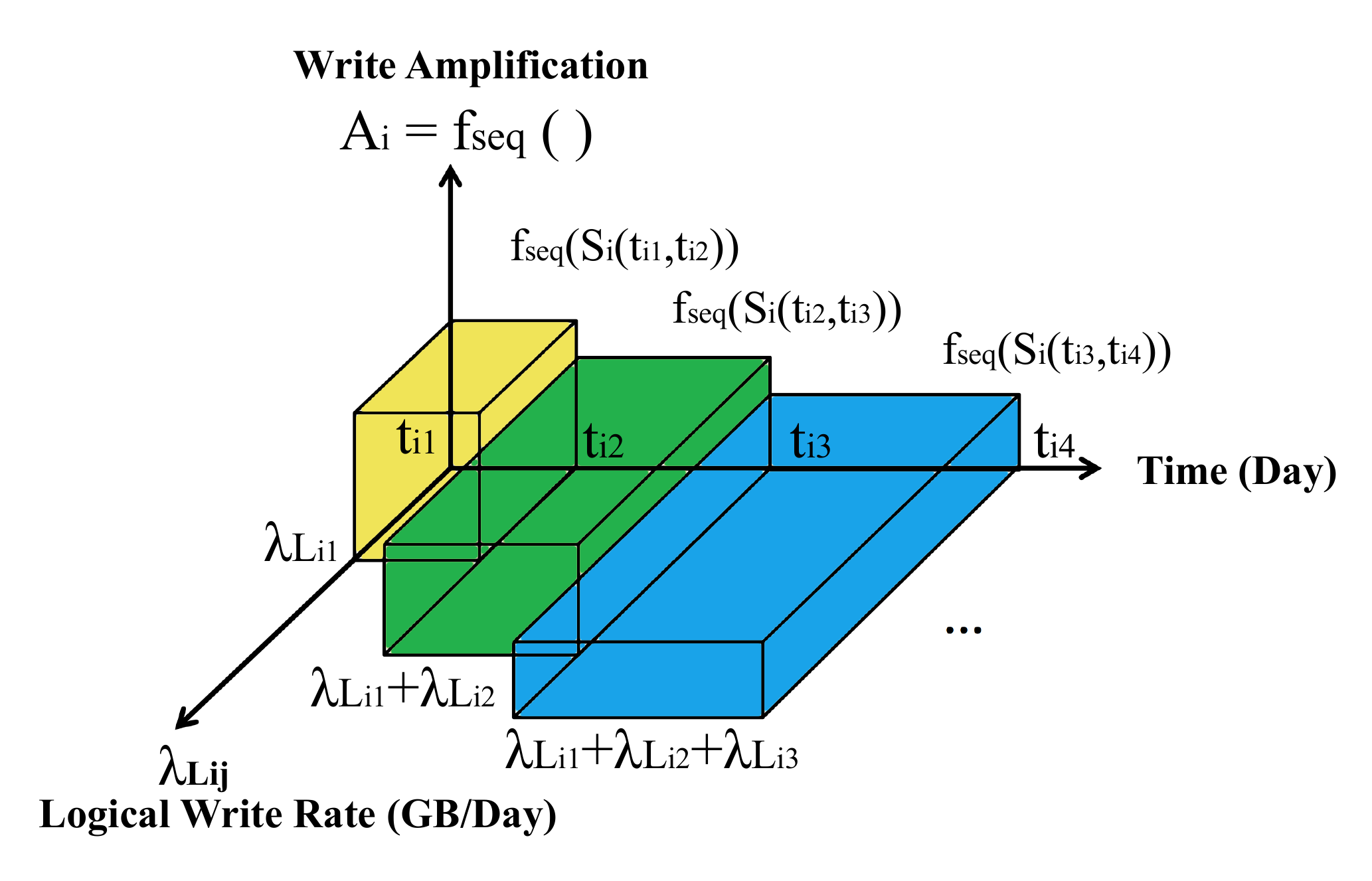}
	\caption{\small An example of write worn out count estimation.}
	\label{FIG:TCO_WRITE_COUNT}
\end{figure}
\setlength{\textfloatsep}{10pt plus 1.0pt minus 2.0pt}

\vspace{-1mm}
\subsubsection{\textbf{Write Wornout Count of Disk $w_i$}}
\label{SUBSUBSEC:TCO_CTM_WWC}


The last item we need to estimate is the current physical write count $w_i$ (in Sec.~\ref{SUBSUBSEC:TCO_CTM_EL}) inside each SSD device. It is hard to exactly measure the overall write count of an SSD during its lifetime. 
However, we can estimate the current write count by adding the estimated write counts of
all the workloads over all past epochs. For each epoch, we multiply the total logical write rate by the corresponding WAF to get the physical write rate. By iterating this process for all epochs, we can finally get the total write wornout count for each disk.
Fig.~\ref{FIG:TCO_WRITE_COUNT} shows a simple example of estimating a disk's write wornout 
when there are multiple workloads executing on disk $i$. Each brick represents an epoch, which is bounded by its workloads' allocation times. The volume of all these
bricks gives the total write wornout count ($w_i$) for disk $i$. To calculate $w_i$, we further convert above-mentioned $\lambda_{L_i}$
and 
$S_i$ to the total logical data write rate function and the sequential ratio function during each epoch $[t_{ix}, t_{i(x+1)})$, respectively: 
$\lambda_{ { L }_{ i } }(t_{ ix },t_{ i{(x+1)} })=\sum _{ j\in { \mathbb{J} }_{ i }(t_{ ix },t_{ i{(x+1)} }) }^{  }{ { \lambda  }_{ { L }_{ ij } } }$, and 
${ S }_{ { i } }(t_{ ix },t_{ { i{(x+1)} } })=\frac { \sum _{ j\in { \mathbb{J} }_{ i }(t_{ ix },t_{ { i{(x+1)} } }) }^{  }{ { \lambda  }_{ { L }_{ ij } }S_{ { ij } } }  }{ \sum _{ j\in { \mathbb{J} }_{ i }(t_{ ix },t_{ { i{(x+1)} } })}^{  }{ { \lambda  }_{ { L }_{ ij } } }  }$, 
%
%
%
where $x$ is the number of workloads executing on disk $i$, and $t_{ix}$ is the arrival time of disk $i$'s $x^{th}$ workload.
$\mathbb{J}_i (t_{ix}, t_{i(x+1)})$ is the set of workloads running on disk $i$ during $t_ix$ and $t_{i(x+1)}$ epoch. $\lambda_{ { L }_{ ij }}$ and $S_{ ij }$ are the write rate and sequential ratio of $j^{th}$ workload in $\mathbb{J}_i (t_{ix}, t_{i(x+1)})$. Therefore, wornout $w_i$ can be calculated as: $w_{ { i } }=\sum _{ t_{ ix }\in { { T } }_{ i } }^{  }{ [\lambda _{ L_{ i } }(t_{ ix },t_{ i(x+1) }) } \cdot f_{ seq }(S_{ i }(t_{ ix },t_{ i(x+1) }))\cdot (t_{ i(x+1) }-t_{ ix })]$.
%
Here, $t_{ix}$ is the starting time of disk $i$'s $x^{th}$ epoch. At the sample moment $t_{i(x+1)}$, we assume there are $x$ workloads running on  disk $i$.  $\mathbb{T}_i$ is the set of arrival times of each workload running on disk $i$. The three parts ($\lambda$, WAF and time) match the three axes from Fig.~\ref{FIG:TCO_WRITE_COUNT}, where each brick stands for each epoch, and the total volume of these bricks is the accumulated write count value of that SSD disk.
Therefore, the data-avg TCO rate $TCO'$ in Eq.~\ref{EQ:TCORD} can be calibrated as: 
\vspace{-1mm} 
\small
\begin{eqnarray}
TCO'=\frac { \sum _{ i=1 }^{ N_{ D } }{ [C_{ I_i }+C'_{ M_{ i } }(T_{ W_i  }+\frac { W_i-w_i }{ \lambda _i \cdot A_i} )] }  }{ \sum _{ j=1 }^{ N_{ W } }{ \lambda _{ j } (T_{Lf_{D(j)}}-T_{I_j}) }  }.
\label{EQ:TCO}
\end{eqnarray}
\normalsize

\vspace{-4.5mm}
\section{Algorithm Design}
\label{SEC:TCO_OA}
\vspace{-1mm}

%
\small
\begin{algorithm}[h]
	\small
	\SetAlFnt{\footnotesize}
	%
	\SetKwProg{ProcMain}{Procedure}{}{}
	\SetKwFunction{FuncMain}{minTCO}
	
	\SetKwProg{ProcTCO}{Procedure}{}{}
	\SetKwFunction{FuncTCO}{TCO\_Assign}{}
	
	\SetKwIF{If}{ElseIf}{Else}{if}{then}{else if}{else}{endif}
	
	\ProcMain{\FuncMain{}}
	{
		\For{incoming new workload $J_N$}{
			\For{$i \gets 1$ to $N_D$}{
				$TCO\_List[i]=TCO\_Assign(i,J_N)$\;
				
			}	
			$SelectedDisk=TCO\_List.minValueIndex()$\;
			$Disk[SelectedDisk].addJob(J_N)$\;
		}
		\KwRet\;
	}
	\ProcTCO{\FuncTCO{$i$,$J_N$}}
	{
		\For{$k \gets 1$ to $N_D$}
		{
			$C_I = getCostInit(k)$\;
			$C_M  = getCostMaint(k)$\;
			\If {k == i}
			{
				$T_{W_k}  = {T_{{J_N}}}-T_{I_{k}}$\;
				$T_{E_k}  = getExpFutureWorkTime(k,J_N) $\; 
				$Data   = getTotalLogWriteAmt(T_{W_k}+T_{E_k})+(T_{W_k}+T_{E_k}-{T_{{J_N}}})*\lambda_{J_N} $\;
			}
			\Else
			{
				$T_{W_k}  = {T_{R_{k}}}  -  T_{I_{k}}$\;
				$T_{E_k}  = getExpFutureWorkTime()$\;
				$Data   = getTotalLogWriteAmt(T_{W_k}+T_{E_k})$\;
			}       
			
			$ TCO    += C_I + C_M * ( T_{W_k} + T_{E_k} )$\;
			$ TotalData += Data$\;            
		}

		\KwRet {TCO/TotalData}\;
	}
	\caption{minTCO}\label{ALG:MINTCO}
\end{algorithm}
\normalsize

Based on the proposed TCO model, we further design a set of online allocation algorithms, called ``{\mf}'', which adaptively allocate new workloads to SSDs in the disk pool. 
The main goal is to minimize the data-avg TCO rate ($TCO'$) of the storage pool and also to conduct disk-level wear leveling during workload deployment and allocation.



\vspace{-3mm}
\subsection{Baseline minTCO}
\label{SUBSEC:TCO_OA_BS}
\vspace{-1mm}

The main functionality of the baseline version of {\mf} is presented in Alg.~\ref{ALG:MINTCO}. 
When a new workload arrives, {\mf} calculates the data-avg TCO rate for the entire disk pool, and then allocates the workload to the SSD that makes the lowest data-avg TCO rate of the entire disk pool. 
Specifically, there are two cases during the calculation of expected lifetime and total logical write amount. The first case is that when a new workload is assigned to disk $k$, we use this new workload's arrival time as the boundary between $T_{W_k}$ and $T_{E_k}$ phases, as shown in Alg.~\ref{ALG:MINTCO} lines 13 to 15 and Fig.~\ref{FIG:TCO_TIME}(c). 
The second case is that when the new workload is not assigned to disk $k$ (Alg.~\ref{ALG:MINTCO} lines 17 to 19), we use $T_{R_k}$ (the arrival time of the most recent workload on disk $k$) as the boundary between $T_{W_k}$ and $T_{E_k}$ phases, as shown in Fig.~\ref{FIG:TCO_TIME}(b). 
As discussed previously, our TCO model is compatible with any WAF models. Here we adopt the WAF model described in Eq.~\ref{EQ:TCO} to implement the functions in Alg.~\ref{ALG:MINTCO} line 14, 15, 18 and 19.
The baseline {\mf} also needs to consider other resource
constraints. 
For example, {\mf} needs to further check if the available spatial (in GB) and throughput (in IOPS) capacities of the chosen SSD are large enough to hold the new workload's working set.
If not, {\mf} moves to the next SSD which has the second lowest data-avg TCO rate.
If no disks have enough capacity, then the workload will be rejected.

\vspace{-3mm}
\subsection{Performance Enhanced minTCO}
\label{SUBSEC:TCO_OA_PE}
\vspace{-1mm}
One limitation of the baseline {\mf} is that it does not balance the load across the SSDs in the pool and thus cannot achieve optimal resources utilization. However, best using of resources (e.g., I/O throughput and disk capacity) is an important goal in real storage system management. To address this limitation, we further develop the performance enhanced manager, namely {\ms}, which considers statistical metrics (i.e., load balancing and resource utilization) as the performance factors in workload allocation.

\vspace{-1mm}
\subsubsection{\textbf{System Resource Utilization}}
\label{SUBSEC:TCO_OA_PE}
\vspace{-1mm}

We consider two types of resources, throughput (IOPS) and space capacity (GB), and calculate the utilization ($U(i,k)$) of disk $i$ when disk $k$ is selected to serve the new workload $J_N$, as:
\small
\begin{eqnarray}
&U(i,k)=\begin{cases} \frac { R_{ U }(i) }{ R(i) } ,&i\neq k 
\\ \frac { R_{ U }(i)+R(J_{ N }) }{ R_{ i } } ,&i=k \end{cases},
\label{EQ:UT}
\end{eqnarray}
\normalsize
where $R_U(i)$,  $R(i)$ and $R(J_N)$ represent the amount of {\em used} resource on disk $k$, the {\em total} amount of resource of disk $i$, and the resource {\em requirement} of workload $J_N$, respectively.
When $i$ is equal to $k$, we have the new requirement (i.e., $R(J_N)$) as extra resources needed on that disk.
This equation can be used to calculate either throughput utilization (i.e., ${U_p{(i,k)}}$) or space capacity utilization (i.e. ${U_s{(i,k)}}$).
The average utilization can be calculated to represent the system utilization of the entire disk pool: $\overline { U(k) } =\frac { \sum _{ k=1 }^{ N_{ D } }{ U(i,k) }  }{ N_{ D } }$.
%
%
Our goal is to increase either average throughput utilization (i.e., ${U_p{(i,k)}}$) or average space utilization (i.e. ${U_s{(i,k)}}$).

\vspace{-1mm}
\subsubsection{\textbf{Load Balancing}}
\label{SUBSEC:TCO_OA_PE}
\vspace{-1mm}
We use coefficient of variation ($CV$) of throughput (or space) utilizations among all disks to ensure the load balancing. 
Specifically, when assigning the workload $J_N$ to disk $k$, we calculate expected $CV(k)$ as: $CV(k)=\frac { \sqrt { \frac { \sum _{ i=1 }^{ N_{ D } }{ [U(i,k)-\overline { U(k) } ]^{ 2 } }  }{ N_{ D } }  }  }{ \overline { U(k) }  }$.
%
%
A smaller $CV(k)$ indicates better load balancing in the datacenter. 
%
%
%

Then, we have our {\ms} algorithm which aims to minimize the data-avg TCO rate
, while achieving best resource utilization 
 and load balancing among disks. 
 {\ms} uses an optimization framework to minimize the objective function under constrains listed in Eq.~\ref{EQ:OBJ}.
%
\vspace{-1mm}
\small
\begin{alignat}{2}\label{EQ:OBJ}
\text{Minimize: }    \nonumber\\   
& f(R_w) \cdot TCO'(k) \nonumber \\  
& - g_s(R_r)\cdot \overline { U_s(k) }  + h_s(R_r)\cdot CV_s(k) \nonumber \\
& - g_p(R_r)\cdot \overline { U_p(k) }  + h_p(R_r)\cdot CV_p(k)  \nonumber \\
\text{Subject to: } \nonumber\\
&   i \in D,  k \in D & \nonumber\\
& 0 \leq TCO'(i,k) \leq {Th}_{c}   &  \nonumber\\
& 0 \leq U_s(i,k) \leq {Th}_{s}   & \nonumber\\
& 0 \leq U_p(i,k) \leq {Th}_{p} 
\end{alignat}
\normalsize
%
Upon the arrival of a new workload $J_N$, we calculate the ``enhanced cost'' of the the disk pool. The object function in Eq.~\ref{EQ:OBJ} contains the TCO rate cost ($f(R_w) \cdot TCO'(k)$), the resource utilization reward ($g_s(R_r)\cdot \overline { U_s(k) }$ and $g_p(R_r)\cdot \overline { U_p(k) }$), and the load unbalancing penalty ($h_s(R_r)\cdot CV_s(k)$ and $h_p(R_r)\cdot CV_p(k)$). 
Notice that $TCO'(k)$ and $TCO'(i,k)$ represent the TCO rate of the entire disk pool and the TCO rate of disk $i$, respectively, when disk $k$ is selected to take the workload.
Non-negative parameters in Eq~\ref{EQ:OBJ} (e.g., $f(R_{w})$, $g_s(R_{r})$, $g_p(R_{r})$, $h_s(R_{r})$ and $h_p(R_{r})$) are the weight functions that are related with the read ratio ($R_r=\frac{readIO\#}{totalIO\#}$) and write ratio ($R_w=\frac{writeIO\#}{totalIO\#}$) of workloads. Finally, the disk with the lowest enhanced cost will be selected for the new workload. The reason behind this is that in the real world, write intensive workloads affect WAF and TCO, and read intensive workloads are sensitive to load balancing degree. In addition, ${Th}_{c}$, ${Th}_{s}$ and ${Th}_{p}$ are used as the upper bounds for TCO, space and throughput resources utilization ratios, respectively.

\vspace{-4mm}
\subsection{RAID Mode minTCO}
\label{SUBSEC:TCO_OA_BS}

In the era of flash drivers, there are also lots of commercial available solutions using RAID mode flash disk arrays in data center to further accelerate the I/O speeds, such as Diff-RAID~\cite{balakrishnan2010differential} and RamSan-500~\cite{hutsell2008depth}. 
RAID technique transforms a number of independent disks into a larger and more reliable logical single entity~\cite{narayanan2009migrating}. 
Motivated by this state-of-the-art trend, we extend {\ms} to support storage systems with RAID mode flash disk arrays. 
The major problem during this transition is that it is almost {\em impossible} to calculate (or even monitor) the exact WAF of RAID flash disk arrays during runtime in real implementation.  
Moreover, for systems with different RAID setups, getting accurate WAF value will be more complicated.  
Therefore, we present an approximate approach that aims to {\em estimate} WAF and TCO to guide the data center manager to make allocation decisions from a long-term datacenter operation of view.

\begin{figure}[ht]
	\centering
	\includegraphics[width=0.32\textwidth]{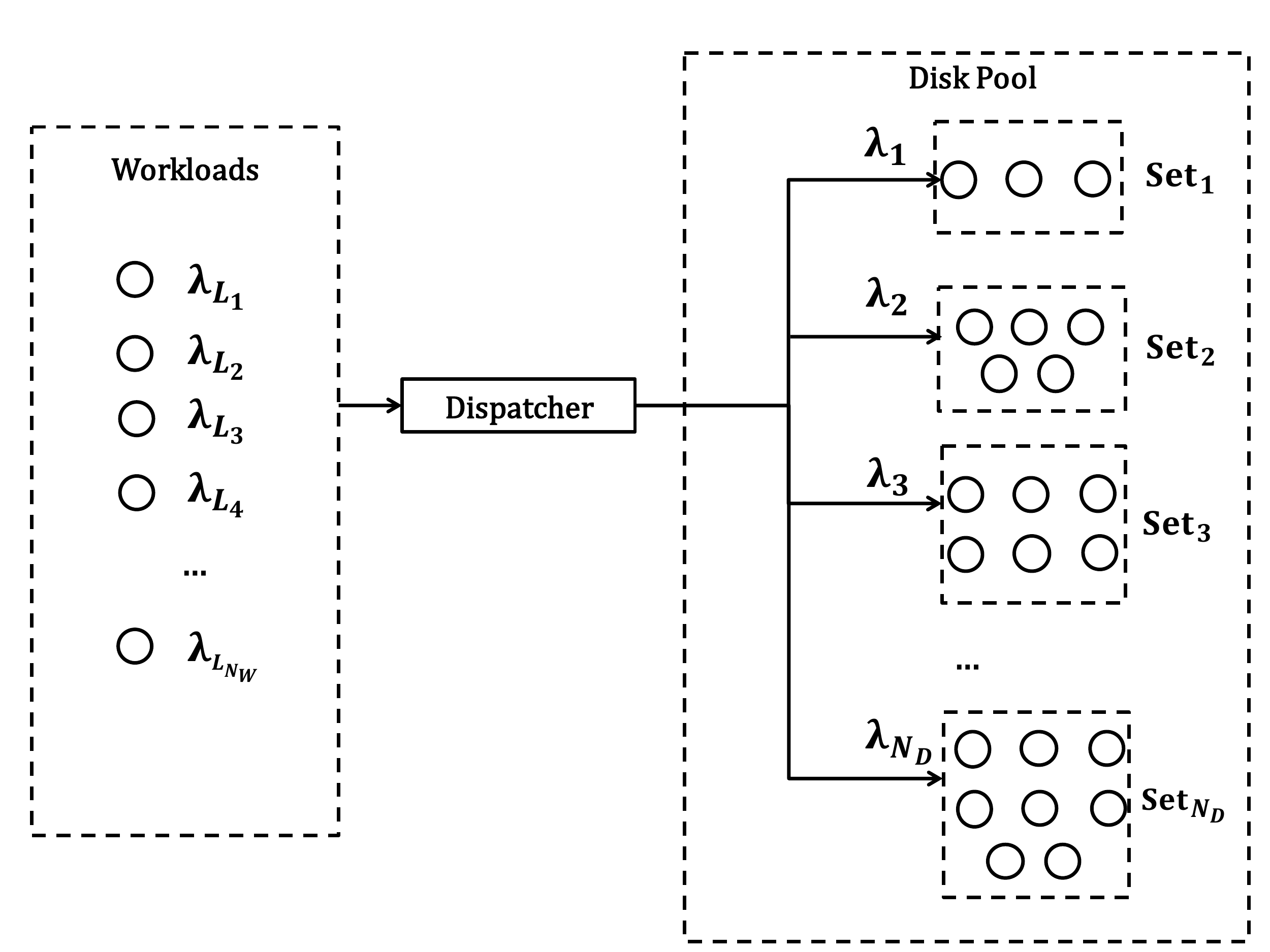}
	\caption{\small Model of RAID data center storage system.}
	\label{FIG:TCO_RAID}
\end{figure}

As shown in Fig.~\ref{FIG:TCO_RAID}, the main idea of our approximation is to treat each  RAID disk array set as a single ``pseudo disk''.
A ``pseudo disk'' can consist of multiple SSDs, which are disks in the same RAID array.
Similar to Fig.~\ref{FIG:TCO_QUEUE}, the ``pseudo disk''  is the minimum unit to be allowed to accept workloads.  
To further simplify the problem, we focus on the cases that disks within each RAID disk set are homogeneous, and different sets can have heterogeneous RAID modes. 
%
%
%
Since other parts in the non-RAID model (i.e., Fig.~\ref{FIG:TCO_QUEUE}) have not been modified in the our RAID model, {\mt} can use the same method to solve the RAID scenario problem by converting TCO and performance metrics of  ``pseudo disks''. 
Specifically, Table~\ref{TAB:RAID} lists the corresponding adjustment parameters for the conversion.

Specifically, some of the factors can be ported from the realm of one disk to a ``pseudo disk'' relatively easily. For example, the costs ($C_I$,$C'_M$) and total write cycle limits ($W$) of a disk set are the sum of those values for each individual disk. 
An N-disk RAID-1 set has CapEx $C_{RAID_I}=N \cdot C_I$, OpEx $C'_{RAID_M}=N \cdot C'_M$, and total write cycle limits $W_{RAID}=N \cdot W$.

However, the estimation of a WAF function for a disk-set is not simple, since WAF is heavily dependent on the actual implementation~\cite{mao2012hpda}. 
From this point on, we focus on a simple implementation which is abstracted for a long-term and large-scale view of the datacenter. 
We observed that the striping parts of RAID-0 and RAID-5 are  following the {\em same I/O  locality and behavior} as the non-striping case, and thus the subset of workload on each disk keeps the same sequential ratio as the non-striped workload~\cite{io1}.
For example, after striping a 100\% sequential stream with <0$\sim$80> pages into four disks (with 10-page striping granularity), the four disks will be assigned pages  <0$\sim$10,41$\sim$50>, <11$\sim$20,51$\sim$60>,  <21$\sim$30,61$\sim$70> and <31$\sim$40,71$\sim$80>, respectively. 
Importantly, those striped subsets of a sequential stream  will be physically written continuously on the disk (e.g., <41$\sim$50> will be physically continuous to <0$\sim$10> on disk 1), therefore the sequentialities will be kept, and thus WAF functions of those striping disks will be the same as that of a single disk~\cite{io2}. 
For the parity disk of RAID-5, we observed that its I/O behavior follows the same locality as the original workload. Therefore, WAFs of RAID-0 and RAID-5 are still the same as a single disk~\cite{raidweb,raidref1,raidref4}.
It is worth mentioning that for an RAID-1 dis sets with 2 or more than 2 disks (we only allow {\em even} number of disks), this RAID-1 set will mirror each I/O to two ``equal'' RAID-0 groups, instead of replicating one single disk to all others which is a huge waste. 
Thus, the WAF function of each (homogeneous) disk set under RAID-1 mode remains the same as that of each individual disk.

Meanwhile, $\lambda_L$ (logical data write rate) and $S$ (spatial capacity) of each disk set also vary due to different RAID modes that might be used across these sets. 
Specifically, RAID-0 stripes the write and does not trigger any additional logical writes to the disk set. 
So, the logical data write rate of the disk set is the same as that of non-RAID mode single disk, 
and the $S$ should be multiplied by $N$. 
RAID-1 mirrors the write (with {\em two} copies), so the logical write rate is doubled compared with the workload's logical write rate. 
%
RAID-5 mode is relatively complex. For each logical write, $N-1$ disks are assigned for striping write, and $1$ disk is left for parity write. Therefore, the overall logical data write rate of the disk set should be scaled by the workload's logical data write rate multiplying $\frac{N}{N-1}$. 

Lastly, RAID-1 and RAID-5 introduce a write penalty ($\rho$)~\cite{raidref2}. 
RAID-1 has two IOPS per write operation, while one RAID-5 write requires four IOPS per write operation~\cite{raidref3}. 
We need to convert an incoming workload's original throughput requirement $P_J$ to its RAID version $P_{RAID_J}$:
\vspace{-1mm} 
\begin{eqnarray}
P_{RAID_J}(i) = P_{J}(i) \cdot R_{W}(i) \cdot \rho(i) + P_{J}(i) \cdot {R_{R}(i)},
\label{EQ:WAF}
\end{eqnarray}
where $\rho(i)$ is the write penalty for disk $i$, which can be obtained from Tab.~\ref{TAB:RAID}. 
For example, a disk set with RAID-1 mode has 4 disks, each of them has $6,000$ IOPS, then the disk set's throughput capacity is $P_{RAID}(i)=6,000 \times 4= 24,000$ IOPS. 
For a new incoming workload which requires 30 IOPS, with write ratio of 40\%, we can obtain its real requirement on RAID-1 mode by as: $P_{RAID_J}=30 \times 40\% \times 2 + 30 \times (1-40\%)=42$ IOPS.
Similarly, if this new coming workload has a logical data write rate of $200$ GB per day, then the equivalent logical data write rate in this RAID-1 disk set is doubled ($200 \times 2 = 400$ GB per day) since it replicates each write I/O.
Furthermore, the actual physical write rate can be estimated based on equivalent logical data write rate and the corresponding write amplification function of the disk.

\begin{table}[]
\centering
\begin{tabular}{|c|c|c|c|c|c|c|c|c|}
\hline
\multirow{2}{*}{\textbf{RAID}}       &\multirow{2}{*}{\textbf{Mode}} & \multicolumn{5}{c|}{\textbf{TCO}}  & \multicolumn{2}{c|}{\textbf{Perf}} \\ \cline{3-9} 
& &  \textbf{$C_I$}                     & \textbf{$C_M'$}   & \textbf{$W$}      & \textbf{$A$} &   \textbf{$\lambda_L$}           & \textbf{$S$}     & \textbf{$\rho$} \\ \hline
$0$                   & Strip & $N$                       & $N$      & $N$ &   $1$         & $1$                & $N$   & $1$  \\ \hline
$1$                   & Mirror & $N$                       & $N$      & $N$ &   $1$         & $2$                & $N/2$     & $2$    \\ \hline
$5$                   & Pair & $N$                       & $N$      & $N$ &   $1$         & $\frac{N}{N-1}$    & $N-1$   & $4$  \\ \hline
\end{tabular}
\caption{\small Conversion table for different RAID modes.}\label{TAB:RAID}\vspace{-4mm}
\end{table}
\setlength{\textfloatsep}{10pt plus 1.0pt minus 2.0pt}

\subsection{Offline Mode {\mf}}
\label{SEC:OM}

{\mf}, {\ms} and {\mt} are working in the so-called {\em online} scenario where the number and specs of disks in the disk pool are known and fixed, while the system has no foreknowledge about those I/O workloads that are keeping coming and waiting to be assigned to the disk pool.
To achieve a lower TCO rate goal, these three algorithms are following the greedy-based approach to obtain the global best solution by choosing the ``local'' best solution.
In real world, there exists another scenario called ``{\em offline}'' scenario, where the datacenter manager needs to allocate {\em all} known workloads to an undecided disk pool in the beginning. In detail, the manager needs to decide the number and type of disks to run the I/O workloads, and to allocate each workload to disks aiming to get a global best data-avg TCO rate solution.
Unfortunately, we cannot directly use these three online-greedy algorithms to solve this offline problem. 
Although it is on theory possible to convert the offline problem into an NP-hard backpack problem with relatively complicated weight and value functions, to solve this problem in field is yet computationally time-consuming.
For instance, due to the core feature of modern datacenters such as elasticity and scalability, both the numbers of workloads and disks are too ``huge'' for existing solutions like \textit{branch and bound method}~\cite{shih1979branch}  or \textit{multiple-objective genetic local search}~\cite{jaszkiewicz2002performance}.

Motivated by this challenge, in this section, we develop a less-costly allocation algorithm {\mff}, which works in  datacenters consisting of \textit{homogeneous} disks. This algorithm starts from rethinking the problem from the relationship between sequential ratio, logical write rate and TCO rate.
In Appendix 2, we prove that sorting all workloads by their sequential ratio and sending them to each disk in the order of sequential ratio is the best solution under ideal conditions. However, in real implementation, to avoid the case that purely based on sequential ratio may lead to capacity or I/O throughput unbalance, we need to manually set up two (or more than two) disk zones and then balance the write rate inside each zone.

\small
\begin{algorithm}
	\small
	\SetAlFnt{\footnotesize}
	
	\SetKwProg{ProcMain}{Procedure}{}{}
	\SetKwFunction{FuncMain}{minTCO-Offline}

	\SetKwProg{ProcTCO}{Procedure}{}{}
	\SetKwFunction{FuncTCO}{Distribute}{}
	
	\SetKwIF{If}{ElseIf}{Else}{if}{then}{else if}{else}{endif}

	\ProcMain{\FuncMain{$\mathbb{J}$,$\mathbb{D}$}}
	{
		\For{each workload $J\in\mathbb{J}$}
		{
			\If{$S_J \geq \varepsilon$}
			{
				$\mathbb{J}_H.add(J)$\;	
				$\lambda_H+=\lambda_J$\;	
			}	
			\Else
			{
				$\mathbb{J}_L.add(J)$\;
				$\lambda_L+=\lambda_J$\;
			}		
		}
		\If {$\left| \frac{\lambda_H-\lambda_L}{\lambda_H+\lambda_L} \right| \geq \delta$}
		{
			/* Greedy Approach */\\
			Distribute($\mathbb{J}$,$\mathbb{D}$)\;
		}
		\Else
		{
			/* Grouping Approach */\\
			$\mathbb{J}_H=\mathbb{J}_H.descendingSortBySeqRatio()$\;  
			$\mathbb{J}_L=\mathbb{J}_L.descendingSortBySeqRatio()$\;  
			Distribute($\mathbb{J}_H$,$\mathbb{D}_H$)\;
			Distribute($\mathbb{J}_L$,$\mathbb{D}_L$)\; 
			
		}
		\KwRet\;
	}

	\ProcTCO{\FuncTCO{$\mathbb{J}$,$\mathbb{D}$}}
	{
		\For{each workload $J \in \mathbb{J}$}
		{
			\If {$C<C_J$ or $P<P_J$}
			{
				\KwRet {Error: "Even empty disk cannot run this workload."}\;
				
			}
			\ElseIf{$size(\mathbb{D})==0$}
			{
				$\mathbb{D}.addNewDisk().addJob(J)$\;
			}
			\Else
			{
				\For {each disk $d \in \mathbb{D}$}
				{
					\If {$remC_d<C_J$ or $remP_d<P_J$}
					{
						$CV[d]=\infty$\;
					}
					\Else
					{
						$updateWriteRateCV(CV,d,\lambda_J)$\;
					}
				}
				
				diskID=minArg(CV)\;
				\If {$CV[diskID]=\infty$}
				{
					/* No  $d\in\mathbb{D}$ can run this workload. */\\
					$\mathbb{D}.addNewDisk().addJob(J)$\;
				}
				\Else	
				{
					$\mathbb{D}[diskID].addJob(J)$\;
				}

			}
		}
		\KwRet\;
	}
	\caption{OfflineDeploy}\label{ALG:MINTCO_OFFLINE}
\end{algorithm}
\normalsize



As shown in Alg.~\ref{ALG:MINTCO_OFFLINE} line 3 to 8, {\mff} first groups the all workloads ($\mathbb{J}$) into high and low sequential ratio groups ($\mathbb{J}_H$ and $\mathbb{J}_L$) by comparing each workload's sequential ratio with a preset threshold $\varepsilon$. 
In fact, {\mff} can be extended to support more than two workload groups and disk zones simply by using a more fine-grained sequential ratio threshold vector $\overrightarrow { \varepsilon } $ to replace the single value threshold $\varepsilon$.
Meanwhile, {\mff} also calculates the total write rate of each group. If write rates of these two zones are close to each other (by comparing their normalized difference with the approach switching threshold $\delta$ in line 9), the algorithm chooses the grouping approach to select a disk among {\em all} disks in the array to allocate the workload. 
Otherwise, the algorithm selects the greedy approach to allocate the workload in a certain disk group (e.g., either the high or low sequential ratio disk zone in our implementation).
Lines 10 and 11 show the greedy approach, which simply calls the distribute function. 
Specifically, the distribute function first checks whether this workload's spatial and throughput capacity requirements exceeds a brand new (empty) disk. If yes, then this workload will be rejected, see in line 21 to 22, where $C$ and $P$ (resp. $C_J$ and $P_J$) are total spatial and I/O throughput capacity of an empty disk (resp. of the current workload), respectively.
Otherwise, it iterates to calculate the CV of write rates among the disk pool if this workload is added to each disk (line 26 to 30, where $remC_d$ and $remP_d$ are remaining spatial and throughput capacity of a certain disk $d$). It attempts to balance the write rate of each disk in the disk zone. 
It finally selects the disk that will result in a minimum CV of write rates (line 31).
Similarly, for the grouping approach, {\mff} sorts each workload group by sequential ratio in the deceasing order (line 14 and 15), and then calls the distribute function to allocate these two groups of workloads to their corresponding zones, i.e., $\mathbb{D}_H$ and $\mathbb{D}_L$ as high and low sequential ratio disk zones, respectively (line 16 and 17). 


\section{Evaluation}
\label{SEC:EV}

\begin{figure*}
	\centering		
	\includegraphics[width=0.98\textwidth]{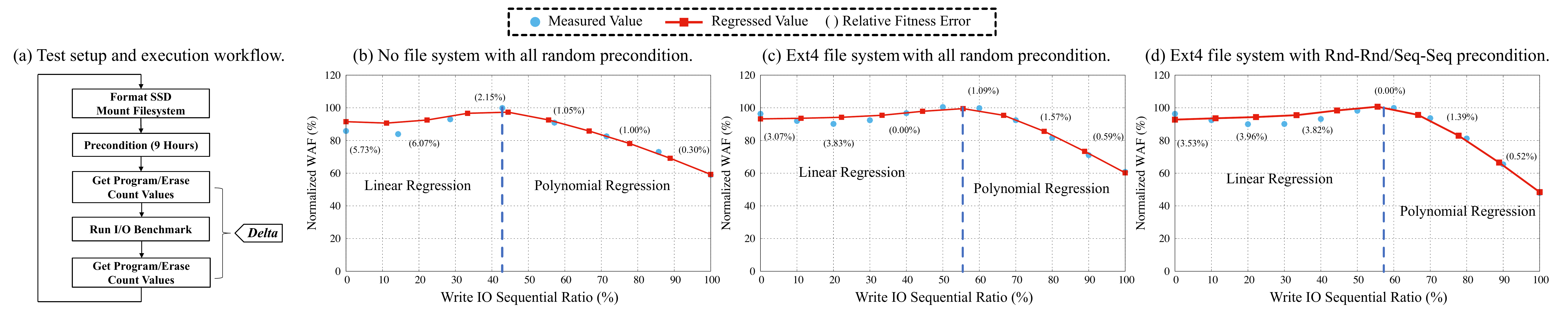}	
	\caption{\small WAF measurement experiment results.}\label{FIG:EV_WAF}
\end{figure*}
\setlength{\textfloatsep}{10pt plus 1.0pt minus 2.0pt}

\subsection{Write Amplification Measurement \& Modeling}
\label{SUBSEC:EV_WAMM}


\subsubsection{\textbf{Hardware Testbed}}
\label{SUBSUBSEC:EV_WAMM_HT}
Most SSD vendors do not provide APIs or performance counters to measure this physical write quantity. Hence, many prior studies (e.g.,\cite{peter,bux2010performance}) have tried to develop models for \textit{estimating} the WAF of an SSD based on a certain criterion.
In this paper, we propose to leverage the data  \textit{directly measured} from SSDs to calculate a WAF function for an SSD. Our goal is to characterize the effects of write traffic from multiple workloads on the WAF, and see if such a characterization can be generalized as a mathematical model. 
Tab.~\ref{TAB:EV_SPEC} and Fig.~\ref{FIG:EV_WAF}(a) shows the testbed specification and the WAF measurement workflow, receptively. 
%

\begin{table}[h]
	\footnotesize
	\center
	\begin{tabular}{|c|c|}
		\hline
		\textbf{Component} & \textbf{Specs}  \\ \hline 
		Processor & Xeon E5-2690, 2.9GHz  \\ \hline
		Processor Cores & Dual Socket-8 Cores       \\ \hline
		Memory Capacity   & 64 GB ECC DDR3 R-DIMMs                     \\ \hline
		Memory Bandwidth  & 102.4$GB/s$  \\ \hline      
		RAID Controller   & LSI SAS 2008  \\ \hline             
		Network           & 10 Gigabit Ethernet NIC                    \\ \hline
		Operating system  & Ubuntu 12.04.5                             \\ \hline
		Linux Kernel      & 3.14 Mainline                              \\ \hline
		FIO Version       & 2.1.10 run with direct I/O                 \\ \hline
		Storage Type		  & NVMe SSD  (Released in 2014)   \\ \hline 	
		Storage Capacity  & 1.6 TB                                  \\ \hline
	\end{tabular}		\caption{\small Server node configuration. }	\label{TAB:EV_SPEC}
\end{table}

\vspace{-6mm} 
\subsubsection{\textbf{Filesystem}}
\label{SUBSUBSEC:EV_WAMM_FS}
We test two representative scenarios: ``formatting with no file system'' and ``formatting with Ext4 file system''.
(1) ``No file system'' mimics the use case like a swap partition, where avoiding a filesystem mainly has three advantages: making more of the disk usable, since a filesystem always has some bookkeeping overhead; making the disks more easily compatible with other systems (e.g., the {\em tar} file format is the same on all systems, while filesystems are different on most systems.); and enabling enterprise user to deploy customized block manager running on the hypervisor without mounting a traditional filesystem.
(2) ``Ext4 file system'' is a very solid file system which has been widely used 
for SSDs in datacenters using linux distributions for the past few years. The journaling that comes with Ext4 is a very important feature for system crash recovery although it causes some acceptable write activity.

\vspace{-2mm} 
\subsubsection{\textbf{Precondition}}
\label{SUBSUBSEC:EV_WAMM_PC}
\small
\begin{table}[]
	\footnotesize
	\centering
	\caption{\small I/O Setups of Preconditioning.}
	\label{TAB:EV_IO}
	\begin{tabular}{|c|c|c|}
		\hline
		& \textbf{Precon. Seq Fill} & \textbf{Precon. Rand Fill}  \\ \hline
		RW         & Write      & Write   \\ \hline
		IODepth    & 16         & 32      \\ \hline
		BlockSize  & 1MB        & 4KB     \\ \hline
		Job Number & 1          & 4       \\ \hline
	\end{tabular}
\end{table}
\normalsize
In order to ensure the SSD in the same state and stimulate the drive to the same performance state at the beginning of each measurement, we also conduct a 9 hour ``preconditioning process''.  
Tab.~\ref{TAB:EV_IO} shows the detail of our preconditioning setups.
In detail, we have the following operations:
``Sequential precondition'' is that, between each measurement, the SSD is completely fulfilled with sequentially I/Os so that all write I/Os in the measurement workloads are overwrite operations, and WAF results will not be independent on garbage collection.
``Random precondition'' will further conduct an additional complete overwrite to the device with random I/Os with 4KB granularity after the sequential preconditioning process to randomize the workset distribution.
``Rnd-Rnd/Seq-Seq precondition'' is the policy that we use the random and sequential precondition for non-100\% sequential and 100\% sequential I/O workloads, respectively. We attempt to use these workloads to observe the ideal write performance (i.e., steady write performance). These two precondition operations can help us simulate different scenarios.
\vspace{-2mm} 

\subsubsection{\textbf{I/O Workloads}}
\label{SUBSUBSEC:EV_WAMM_IW}

\begin{table}[ht]
	\centering
	\caption{\small Statistics for {\em {part}} of I/O workloads we use. }\label{TAB:EV_TRACE}
	\begin{tabular}{|c|c|c|c|c|c|c|}
		\hline
		Trace       &   $S$     &  $\lambda$& $P_{Pk}$  & 	$R_W$  &   $WSs$\\ 
		Name	    & 	($\%$)  &  	($GB/day$)&	 ($IOPS$) &	($\%$)    &	($GB$)  \\ \hline
		mds0		&   31.52	&	21.04	&	207.02	&	88.11	&	6.43\\ \hline
		prn0		&   39.13	&	131.33	&	254.55	&	89.21	&	32.74\\ \hline
		proj3		&   72.06	&	7.50    &	345.52	&	5.18	&	14.35\\ \hline
		stg0		&   35.92	&	43.11	&	187.01	&	84.81	&	13.21\\ \hline 
		usr0		&   28.06	&	37.36	&	138.28	&	59.58	&	7.49\\ \hline 
		usr2		&   46.10	&	75.63	&	584.50	&	18.87	&	763.12\\ \hline
		wdv0		&   30.78	&	20.42	&	55.84	&	79.92	&	3.18\\ \hline
		web0		&   34.56	&	33.35	&	249.67	&	70.12	&	14.91\\ \hline \hline
		hm1		&	25.15	&	139.40	&	298.33	&	90.45	&	20.16\\ \hline
		hm2		&	10.20	&	73.12	&	77.52	&	98.53	&	2.28\\ \hline
		hm3		&	10.21	&	86.28	&	76.11	&	99.86	&	1.74\\ \hline
		onl2		&	74.41	&	15.01	&	292.69	&	64.25	&	3.44\\ \hline \hline
		Fin1		&	35.92	&	575.94	&	218.59	&	76.84	&	1.08\\ \hline
		Fin2		&	24.13	&	76.60	&	159.94	&	17.65	&	1.11\\ \hline
		Web1		&	 7.46	&	0.95	&	355.38	&	0.02	&	18.37\\ \hline
		Web3		&	69.70	&	0.18	&	245.09	&	0.03	&	19.21\\ \hline
	\end{tabular}
\end{table}

In order to study the effects of sequential traffic on WAF, we conduct an experiment that can control the amount of sequential traffic being sent to an SSD. 
Most workloads in real systems are a certain mixture of sequential and random I/Os.
To mimic such real situations, we generate mixed workloads by using an I/O testing tool FIO~\cite{fio}. 
We also make changes to an 1.6TB NVMe SSD firmware to parse the value of (page) program and (block) erase counters.  We investigate the write amplification factor as the ratio of sequential and random accesses, and the changes of these counters, as shown as ``delta'' in Fig.~\ref{FIG:EV_WAF}(a). 


\vspace{-1mm} 
\subsubsection{\textbf{WAF Results and Modeling}}
\label{SUBSUBSEC:EV_WAMM_WRM}
\vspace{-1mm}

%

Fig.~\ref{FIG:EV_WAF}(b)-(d) show three representative cases from our WAF experimental results, which present the normalized WAF as a function of different sequential ratios on write I/Os. The WAF data points are normalized by the largest WAF across different workload sequential ratios (e.g., the WAF under 40.22\% sequential ratio in Fig.~\ref{FIG:EV_WAF} (b)). Thus, the original WAF is $\in [1,+\infty)$, while the normalized WAF is $\in [0,1]$.

First, we can see that WAF curves in the three figures are similar, i.e., the curves can be regressed into  two stages: a flat linear regression stage and a dramatically decreasing polynomial regression stage.
The former part shows that the write amplification factor of mixed workloads is almost identical to that of a pure random workload and keeps almost constant before a turning point (around 40\% to 60\%). But, after this turning point, the WAF dramatically decreases. In other word, a small fraction of random accesses is necessary to intensify the write amplification factor.

We further regress the WAF ($A$) as a piecewise function of sequentiality of I/O operations in the workload as shown in Eq.~\ref{EQ:EV_WAF}
, where $\alpha$, $\beta$, $\gamma$, $\mu$ and $\varepsilon$ are parameters, and $S$ is the sequential ratio. 
\vspace{-2mm}
\small
\begin{eqnarray}
A=f_{ seq }(S)=
\begin{cases}
\alpha S+\beta , &\qquad S\in [0,\varepsilon ] \\
\eta S^{ 2 }+\mu S+\gamma ,&\qquad S\in (\varepsilon ,1]
\end{cases}
\label{EQ:EV_WAF}
\end{eqnarray}
\normalsize
At the turning point $S=\varepsilon$, we have $\alpha \varepsilon +\beta =\eta \varepsilon^{ 2 }+\mu \varepsilon+\gamma$.
Additionally, $\alpha$ is close to zero since the linear regression stage is relatively smooth. We carry out these experiments multiple times on a number of NVMe SSDs, and draw our conclusions as follows.
We believe that such a mathematical model of sequential write traffic vs. WAF can be constructed for most devices, and each SSD can have its own unique version of WAF function, depending on a number of its hardware-related factors  (FTL, wear leveling, over-provisioning etc.). However, being able to regress a mathematical model for the problem forms the basis of the rest of this paper. 
Additionally, we also observe that the regression turning point of the non-filesystem case (Fig.~\ref{FIG:EV_WAF}(b)) comes earlier than Ext4's (Fig.~\ref{FIG:EV_WAF}(c) and (d)). This validates the fact that Ext4's (bookkeeping) overhead is heavier than the raw disk.
Moreover, when the sequential ratio is 100\%, the WAF under ``Rnd-Rnd/Seq-Seq precondition'' case (Fig.~\ref{FIG:EV_WAF}(d)) is lower than that under the ``All-Rnd precondition'' case (Fig.~\ref{FIG:EV_WAF}(c)). This validates that in the former case, the steady write performance can be reached.

\begin{figure*}[ht]
	\centering
	\includegraphics[width=0.93\textwidth]{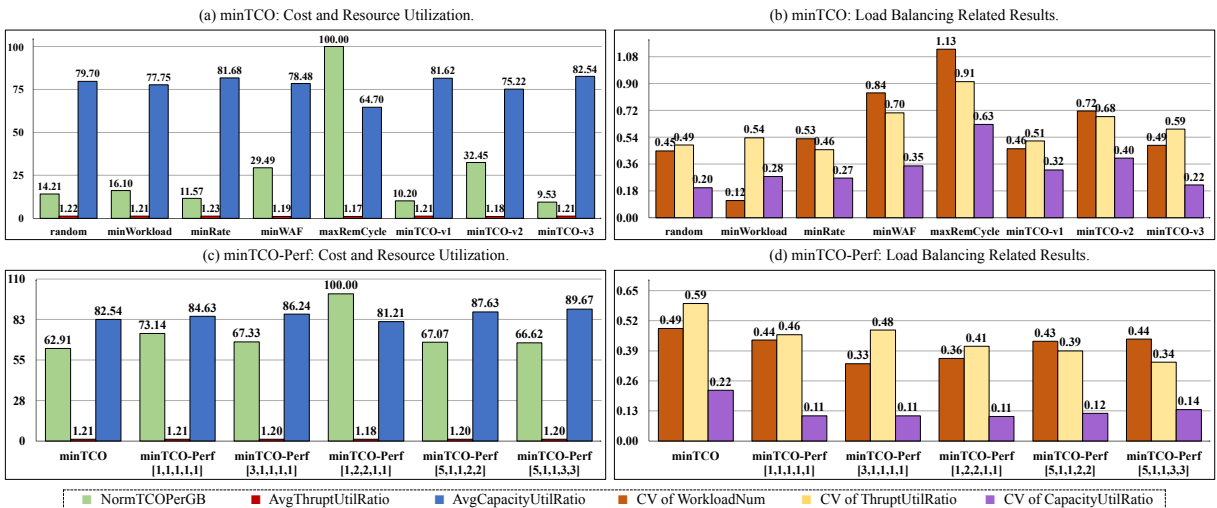}	
	\caption{\small TCO rate, resource utilization, and load balancing results under {\mf} and {\ms}.}  	
	\label{FIG:EV_EX8}	
\end{figure*}
\setlength{\textfloatsep}{10pt plus 1.0pt minus 2.0pt}

\vspace{-2mm} 
\subsection{TCO Evaluation}
\label{SUBSEC:EV_TE}
\subsubsection{\textbf{Benchmarks and Metrics}}
\label{SUBSUBSEC:EV_TE_BM}

In this section, we plug the WAF model regressed from the real experiments to our TCO model, and then evaluate our {\mf} algorithms. Our trace-driven simulation experiments are conducted based on the spec of real NVMe disks 
and enterprise I/O workloads. Specifically, we evaluate more than one hundred enterprise workloads from MSR-Cambridge~\cite{MSR}, FIU~\cite{SNIA} and UMass~\cite{UMASS} trace repositories. These workloads represent applications widely used in real cloud storage systems, such as financial applications, web mail servers, search engines, etc. 

Tab.~\ref{TAB:EV_TRACE} shows the statistics for some of these workloads (out of more than 100 workloads that we are using), where $S$ is the sequential ratio of write I/O (i.e., the ratio between the amount of sequential write I/Os and the amount of total write I/Os), $\lambda$ is the daily logical write rate (GB/day), $P_{Pk}$ is the peak throughput demand with the $5min$ statistical analyses window, $R_W$ is the write I/O ratio (i.e., the ratio between the amount of write I/Os and the amount of total I/Os), and $WSs$ is the workings set size (i.e., the spatial capacity demand). 
The arrival process of these workloads is drawn from an exponential distribution. We use the following metrics to evaluate our {\mf} algorithms: (1) cost per GB during the expected lifetime: the total logical data-averaged TCO during the expected lifetime ($TCO'_{LfPerData}$); 
(2) resource utilization: the average throughput and space capacity utilization ratios among all disks; 
and (3) load balancing: the $CV$ of resource utilization ratio across all disks.


\vspace{-3mm}
\subsubsection{\textbf{TCO Experimental Results}}
\label{SUBSUBSEC:EV_TE_TER}

We implement both baseline {\mf} and the performance enhanced {\ms}. Additionally, three versions of {\mf} are considered, such that {\mf}-$v1$ uses the TCO of expected lifetime ($\sum _{ i=1 }^{ N_{ D } }{ (C_{ I_{ i } }+C '_{ M_{ i } }\cdot T_{ Lf_{ i } }) }$),  {\mf}-$v2$ uses the TCO model of expected lifetime per day ($\frac { \sum _{ i=1 }^{ N_{ D } }{ (C_{ I_i }+C' _{ M_{ i } } \cdot T_{{Lf}_i}) }  }{ \sum _{ i=1 }^{ N_{ D } } T_{ {Lf}_{ i } } }$), 
and {\mf}-$v3$ uses the TCO model of expected lifetime per GB amount  ($\frac { \sum _{ i=1 }^{ N_{ D } }{ (C_{ I_i }+C' _{ M_{ i } } \cdot T_{{Lf}_i}) }  }{ \sum _{ j=1 }^{ N_{ W } }{ D_j }  }$). As expected, none of these baseline {\mf} algorithms consider load balancing and resource utilization during allocation. 
For comparison, we also implement other widely used allocation algorithms, including $maxRemCycle$ which selects the disk with the greatest number of remaining write cycles, $minWAF$ which chooses the disk with the lowest estimated WAF value \textit{after} adding the newly incoming workload, $minRate$ which chooses the disk with the smallest sum of all its workloads' logical write rates, and $minWorkloadNum$ which selects the disk with the smallest number of workloads.
%

\begin{figure*}[ht]
	\centering
	\includegraphics[width=0.96\textwidth]{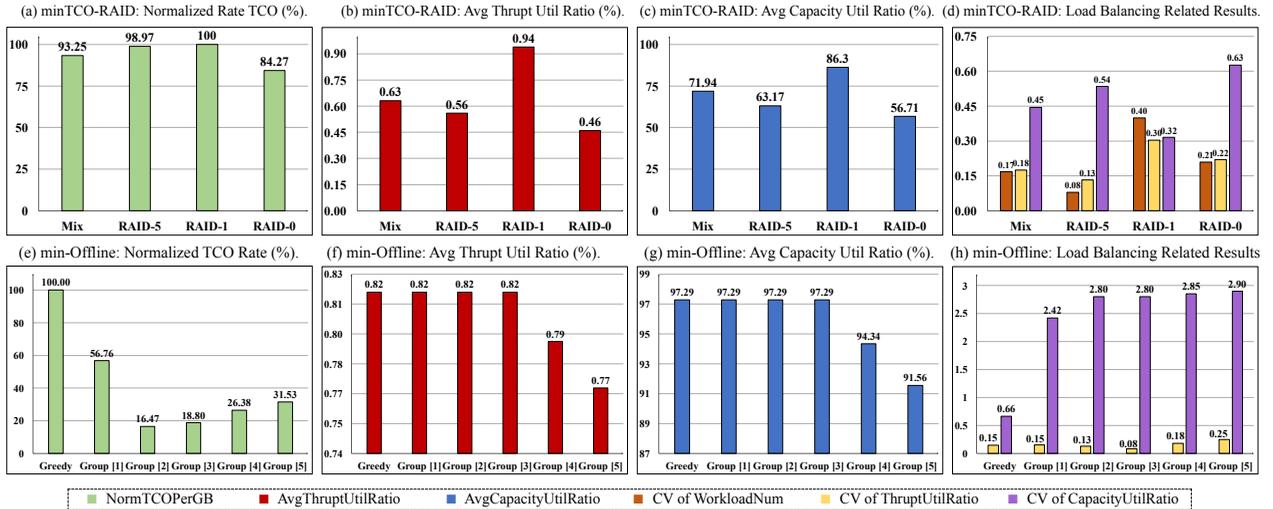}	
	\caption{\small TCO rate, resource utilization, and load balancing results under {\mt} and {\mff}.}  	
	\label{FIG:EV_OFFLINE}	
\end{figure*}

\noindent \textbf{\textit{(1) minTCO}}

We first conduct the experiments running on the disk pool which consists of 20 disks, with 9 different models of NVMe SSDs (available on market in fall 2015). 
In our implementation, we mix about 100 workloads from MSR, FIU, and Umass with exponentially distributed arrival times in 525 days. 
Fig.~\ref{FIG:EV_EX8}(a) and (c) show the results of data-avg TCO rates and resource (I/O throughput and space capacity) utilizations under different allocation algorithms. Fig.~\ref{FIG:EV_EX8}(b) and (d) further present the performance of load balancing, e.g., $CV$s of workload number and resource utilizations.
First, as shown in Fig.~\ref{FIG:EV_EX8}(a) and (c), {\mf}-v3 achieves the lowest data-avg TCO rate(\$/GB).
We also observe that among the {\mf} family, {\mf}-$v2$ performs the worst, see Fig.~\ref{FIG:EV_EX8}(a), and obtains the largest $CV$s of allocated workload numbers. The reason is that, to some extent, {\mf}-$v2$ aims at maximizing the expected life time by sending almost all workloads to a single disk, in order to avoid ``damaging'' disks and increasing the TCO.
Therefore, it cannot ``evenly'' allocate the workloads.
We further find that $maxRemCycle$ performs the worst among all allocation algorithms, because it does not consider the TCO as well as the varying WAF 
due to different sequentialities of the running workloads. 
In summary, $minTCO$-$v3$ is the best choice which considers expected life time, cost and expected logical data amount that can be written to each disk.

\noindent \textbf{\textit{(2) minTCO-Perf}}

We next implement {\ms} which is based on {\mf}-$v3$, and considers the data-avg TCO rate as the criterion to choose the disk for the new workload. As described in Sec.~\ref{SUBSEC:TCO_OA_PE}, {\ms} uses Eq.~\ref{EQ:OBJ} to find the best candidate under the goal of minimizing TCO, maximizing resource utilization, and balancing the load. There are a set of weight functions (i.e., $f(R_{w})$, $g_s(R_{r})$, $g_p(R_{r})$, $h_s(R_{r})$ and $h_p(R_{r})$) used in Eq.~\ref{EQ:OBJ} .
To investigate the effects of these weight functions, we conduct sensitivity analysis on the average values for the five weight functions in Eq.~\ref{EQ:OBJ}.
After trying different approaches, and choose the linear function approach to implement weight functions. We show the over-time average value of each function normalized by the minimum function one. For example, ``[5,1,1,2,2]'' means that all values are normalized by the second weight function ($g_s(R_{r})$).
In Fig.~\ref{FIG:EV_EX8}(c) and (g), we observe that space capacity (instead of I/O throughput) is always the system bottleneck (i.e., with high utilization) across different approaches. This is because NVMe SSDs 
support up to $64K$ I/O queues and up to $64K$ commands per queue (i.e., an aggregation of $2K$ MSI-X interrupts). 
Meanwhile, workloads we are using here are collected from traditional enterprise servers, which have not been optimized for NVMe's revolutionary throughput capacity improvement.
We also find that with a slight sacrifice in TCO, {\ms} can improve both resource utilization and load balancing.
Fig.~\ref{FIG:EV_EX8}(c) further show that ``[5,1,1,3,3]'' is the best choice among all cases, which is $3.71\%$ more expensive than the baseline $minTCO$, but increases the space utilization ratio by $7.13\%$, and reduces $CV$ of throughput and space capacity utilization by $0.25$ and $0.8$, respectively. This is because {\ms} sets TCO and space capacity higher priorities.


\noindent \textbf{\textit{(3) minTCO-RAID}}

Our third experiment is for different RAID modes. We group 6 same-model NVMe disks into a set (internal homogeneous), but allow various sets to have different models (external heterogeneous).
In total, we have 8 sets with 48 disks. 
Fig.~\ref{FIG:EV_OFFLINE}(a) to (d) compare results of RAID-0 striping, RAID-1 mirroring, RAID-5 pairing and mix of these three modes under {\mt}. RAID-1 has the highest TCO since it duplicates each I/O, followed by RAID-5 which has $1/N$ space used for replica.  RAID-0 has zero replica so its TCO per data is the lowest. The results of the mix case is in between RAID-1 and RAID-5. No doubt that the average utilization ratios of both throughput and capacity of RAID-1 are the greatest among all, since it mirrors each I/O. 
Since we assign a larger weight for spatial capacity in the objective function than others, RAID-1, as the most spatial-capacity-sensitive RAID mode, obtains the most benefit.
\begin{figure*}[ht]
	\centering	
	\includegraphics[width=0.95\textwidth]{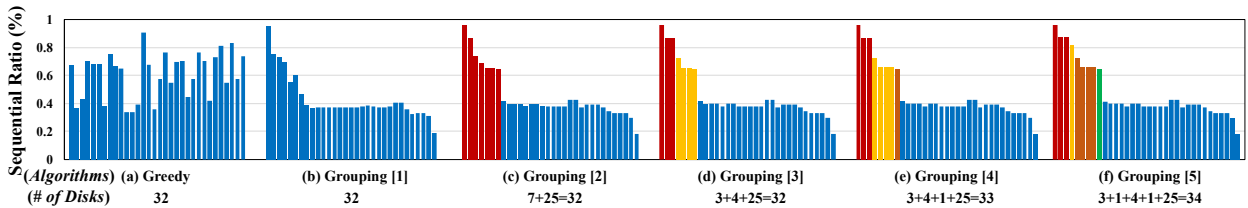}	
	\caption{\small Sequential ratios and disk-zone distributions under different approaches.}  \label{FIG:EV_OFFLINE_DIST}	
\end{figure*}
\setlength{\textfloatsep}{10pt plus 1.0pt minus 2.0pt}

\noindent \textbf{\textit{(4) minTCO-Offline}}

The last experiment is to investigate the performance of {\mff} in the offline scenario. We attempt to allocate 1359 MRS, FIU and Umass I/O workloads to a datacenter with homogeneous disks.
Specifically, {\mff} needs to decide how many disks we need and how to allocate workloads to the disk pool to minimized the data-avg TCO rate. 
In our implementation, we build both the greedy and grouping allocators to evaluate the performance as well as to tune the switching threshold $\delta$.
\begin{figure}[h]
	\centering	
	\includegraphics[width=0.38\textwidth]{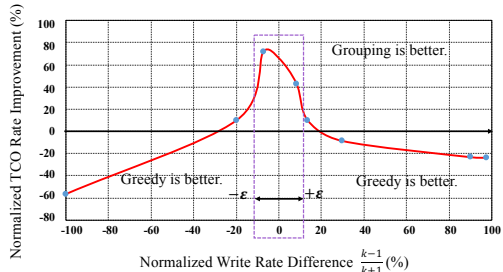}	
	\caption{\small Validation on approach switching threshold.} \label{FIG:EV_OFFLINE_TURNING}	
\end{figure}
According to our rationales in Sec.~\ref{SEC:OM}, the best use case for the grouping approach is to sort and allocate workloads to disks by their sequential ratios while keeping write rates of each disk similar.  In other words, the number of zones should be the same as the number of disks. But in real implementation, we need to consider the spatial capacity and I/O throughput constrains, therefore we tune different numbers of zones to investigate the performance.
As shown in Fig.~\ref{FIG:EV_OFFLINE}(e) to (h), we observe that 2-zone grouping approach has the lowest TCO rate; and the greedy and grouping approaches with 1 to 3 zones have almost the same throughput and capacity utilization ratios.  
In terms of load balancing, the 3-zone grouping case's I/O throughput is best balanced, and the greedy approach achieving the best balance in spatial capacity.

To further investigate the reason of performance difference, Fig.~\ref{FIG:EV_OFFLINE_DIST} illustrates the disk-zone distributions and the sequential ratios of each disk under different algorithms.
The greedy approach's sequential ratio distribution is similar to a randomized curve, while,   all these grouping approaches unsurprisingly have monotone decreasing curves. 
Additionally, the larger the number of disk zones is, the better the disk pool is sorted by sequential ratios.
However, the trade-off of allocating more disk zones is that it triggers more ``unnecessary'' disks compared with less disk zones, when the sequential ratio distribution of the entire workload set is not equalikely matching the sequential ratio threshold vector $\vec{ \varepsilon} $ (i.e., some disk zones have very rare workloads).
For example, as shown in Fig.~\ref{FIG:EV_OFFLINE_DIST}, if we divide workloads into more than three disk zones, we need 33 disks (4 zones, see Fig.~\ref{FIG:EV_OFFLINE_DIST}(e)) and 34 disks (5 zones, see Fig.~\ref{FIG:EV_OFFLINE_DIST}(f)), on the other hand, we only need 3 disks when we consider 3 or less zones (see Fig.~\ref{FIG:EV_OFFLINE_DIST}(b) to (d)),  or even the greedy approach (Fig.~\ref{FIG:EV_OFFLINE_DIST}(a)).
The reason is that as indicated in WAF measurement, NVMe SSD's WAF is almost identical and keeps almost constant before a turning point (around 40\% to 60\%), so there is no need to ``fine-grainedly'' divide the sequential ratio range from zero to the turning point.
Therefore, based on these results, we choose the number of zones as 2 since it has the lowest data-avg TCO and relatively good resource utilization and load balance performance. 

Additionally, we further conduct a set of sensitivity analysis on different logical write rates to decide the threshold of switching by using different I/O workloads with different overall sequential ratios.
Fig.~\ref{FIG:EV_OFFLINE_TURNING} shows the normalized TCO rate improvement degree (i.e., $\frac{TCO'(Greedy)-TCO'(Grouping)}{TCO'(Greedy)}$) vs.  normalized write rate difference of two workload groups (i.e., $\frac{\lambda_L-\lambda_H}{\lambda_H+\lambda_L}=\frac{k-1}{k+1}\in[-100\%,100\%]$, where $k\in[0,+\infty)$).
Here we fix the sequential ratio threshold $\epsilon$ to 0.6.
We observe that when $k$ is less than 1.31 (i.e., $\delta=13.46\%$), the grouping approach is better, which validates the ``$k\rightarrow1$'' condition in Sec.~\ref{SEC:OM}.
Therefore, we set {\mff}'s approach-switching threshold to $\delta=13.46\%$.

\vspace{-3mm}
\section{Related Work}
\label{SEC:RW}
Few prior studies that have focused on the long-term costs of SSD-intensive storage systems with SSDs so far, especially in the context of datacenters. Majority of the existing literature investigates SSD-HDD tiering storage systems.
\cite{vfrmg}  was proposed to reduce the cost of a SSD-HDD tiering storage system by increasing both temporal and spatial update granularities. The ``cost'' in vFRM is the I/O latency and delay, rather than price. 
\cite{li2014importance} built a cost model that also considers the lifetime cost of ownership including energy and power costs, replacement cost, and more. They assume that the ``trade-in'' value of the disk is a linear function of its available write cycles.
%
%
%
Our previous work~\cite{mintcopp} also reveals the relationship between WAF and write I/O sequential ratio. 
%
Meanwhile, in terms of budget-driven workload allocation method, \cite{ghandeharizadeh2015memory} recently presents a systematic way to determine the optimal cache configuration given a fixed budget based on frequencies of read and write requests to individual data items.
\cite{jun2015workload} discussed modeling NVMe workload for optimizing TCO. However, it only addressed on the online and per-I/O-request scheduling problem by minimizing the ``cost'' in terms of workloads’ latencies, while our work focus on  both online and offline allocation problem based on the dollar-per-GB TCO model in a long term view.
Write amplification of an SSD depends on the FTL algorithm deployed in the controller. The studies in \cite{fast,bast,kast,last} presented FTL algorithms that are commonly adopted in public domain. However, SSD vendors do not publicly reveal the FTL algorithms to customers due to confidentiality issues. 

\vspace{-5mm}
\section{Conclusion}
\label{SEC:CO}
\vspace{-2mm}
In this paper, we characterize the write amplification (WA) of SSDs as a function of fraction of sequential writes in a workload. We plug this WAF function into our proposed Total Cost of Ownership (TCO) model, which also considers capital and operational cost, estimated lifetime of flash under different workloads, resource restrictions and performance QoS. Based on the TCO model, we build the online workload allocation algorithm {\mf} and {\ms}. Experimental results show that {\mf} reduces the ownership cost, and {\ms} further balances the load among disks and maximize the overall resource utilization, while keeping the TCO as low as possible. 
{\mt} extends the above solution to RAID mode of enterprise storage environment.
Last but not least, {\mff} presents a less-costly solution for the offline allocation scenario to minimize the overall TCO.
%
%

\vspace{-5mm} 

\bibliographystyle{IEEEtran}{\small \bibliography{08_References}}

\newpage
\begin{IEEEbiography}
[{\includegraphics[width=1in,height=1.25in,clip,keepaspectratio]{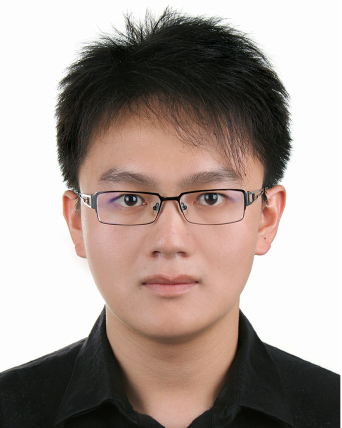}}]
{Zhengyu Yang}
is a Senior Software Engineer at Samsung Semiconductor Inc. He received his Ph.D. degree at Northeastern University, Boston. He graduated from the Hong Kong University of Science and Technology with a M.Sc. degree in Telecommunications, and he obtained his B.Sc. degree in Communication Engineering from Tongji University, Shanghai, China. 
His research interests are cache algorithm, deduplication, cloud computing, datacenter storage and scheduling optimization.
\end{IEEEbiography}

\vspace{-0.5in}
\begin{IEEEbiography}
[{\includegraphics[width=1in,height=1.25in,clip,keepaspectratio]{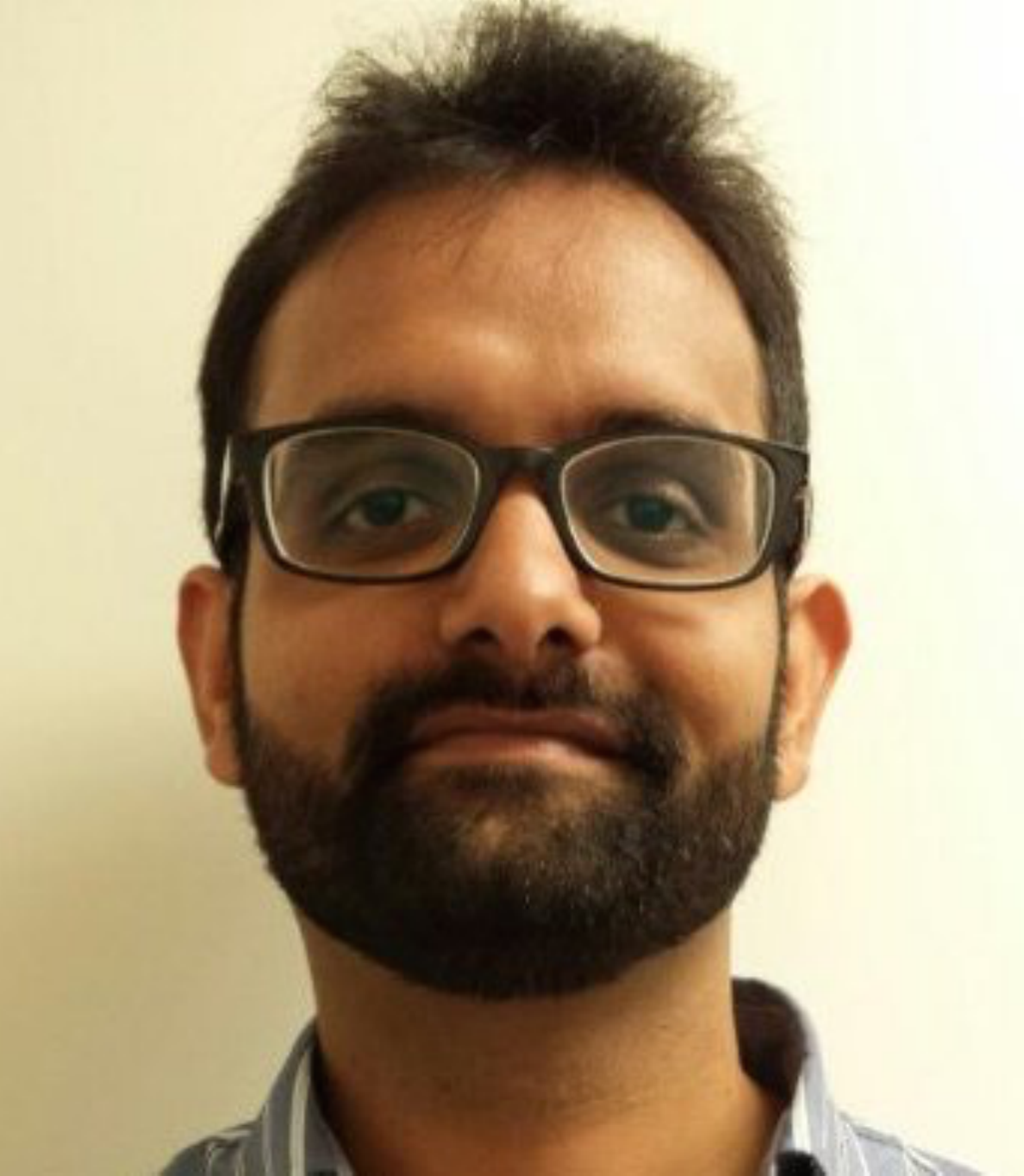}}]{Manu Awasthi}
 is an Associate Professor at Ashoka University, India. During this work, he was a Senior Staff Engineer at Samsung Semiconductor Inc., San Jose, CA. He received his Ph.D. degree in Computer Science from University of Utah, UT in 2011. His research interests are performance evaluation, storage reference architectures, and characterization of datacenter applications.
\end{IEEEbiography}

\vspace{-0.5in}
\begin{IEEEbiography}
	[{\includegraphics[width=1in,height=1.25in,clip,keepaspectratio]{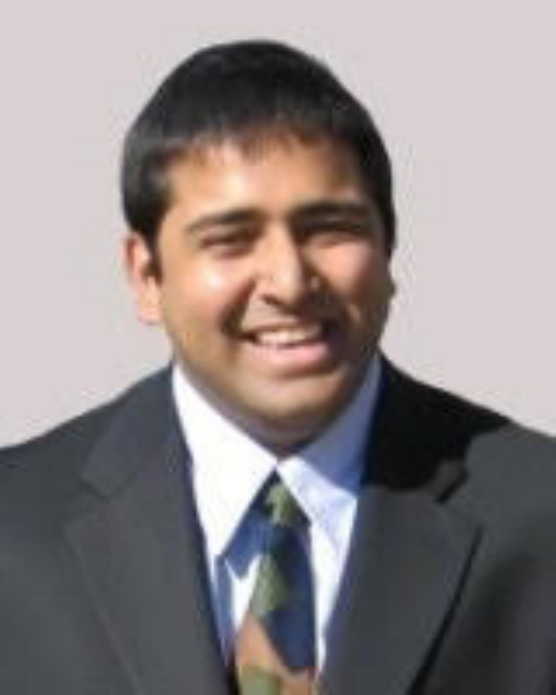}}]
	{Mrinmoy Ghosh}
	is a Performance and Capacity Engineer at Facebook Inc., Menlo Park, CA. During this work, he was a Senior Staff Engineer at Samsung Semiconductor Inc., San Jose, CA. He received his Ph.D. degree in Computer Engineering from Georgia Institute of Technology, GA in 2008. His research interests are characterizing datacenter applications, analyzing storage stack performance, NVMe SSD performance, and remote storage performance.
\end{IEEEbiography}

\vspace{-0.5in}
\begin{IEEEbiography}
	[{\includegraphics[width=1in,height=1.25in,clip,keepaspectratio]{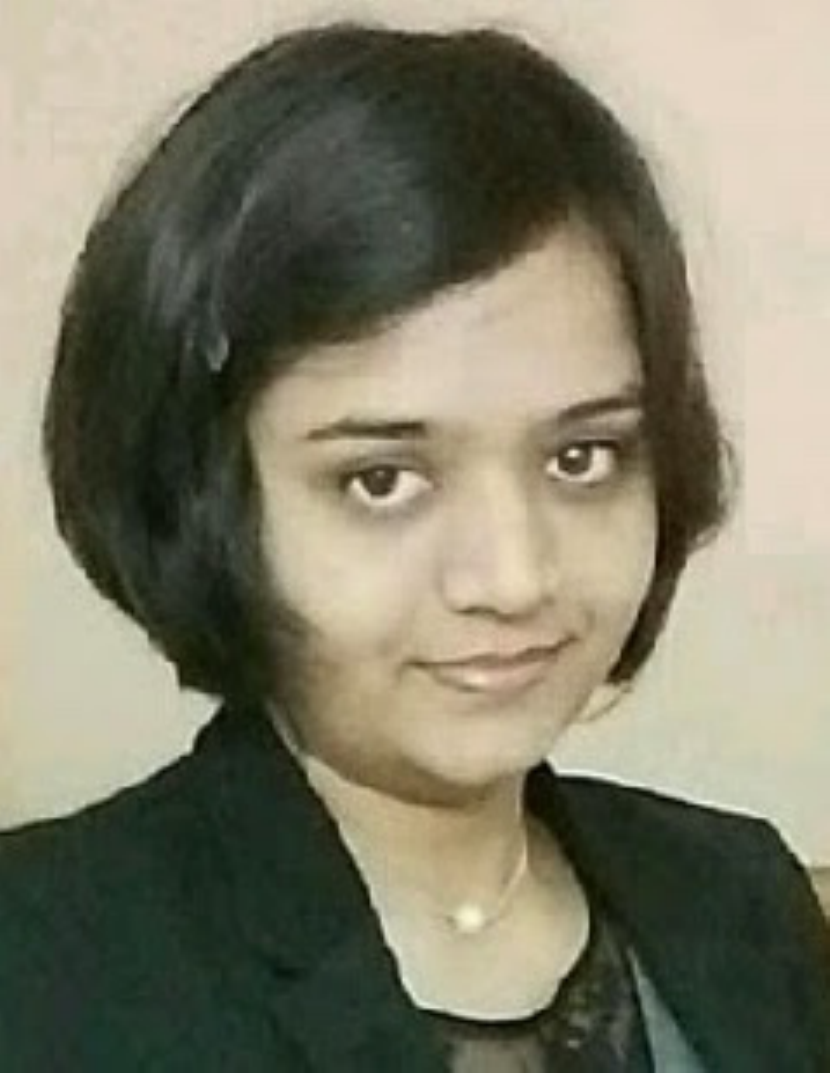}}]
	{Janki Bhimani}
	is a Ph.D. candidate working with Prof. Ningfang Mi at Northeastern University, Boston. Her current research focuses on performance prediction and capacity planning for parallel computing heterogeneous platforms and backend storage. She received her M.S. from Northeastern University in Computer Engineering. She received her B.Tech. from Gitam University, India in Electrical and Electronics Engineering.
\end{IEEEbiography}

\vspace{-0.5in}
\begin{IEEEbiography}
[{\includegraphics[width=1in,height=1.25in,clip,keepaspectratio]{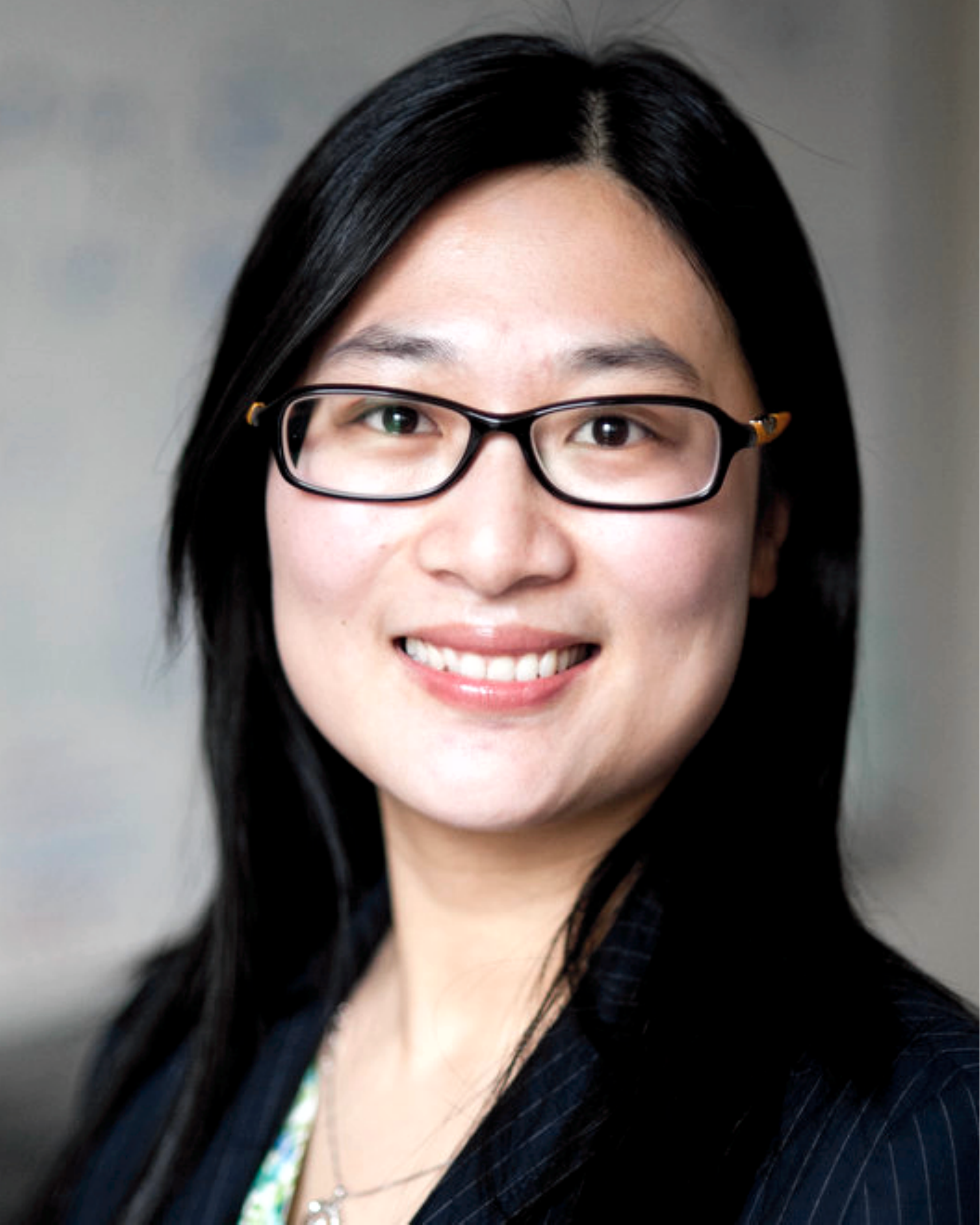}}]{Ningfang Mi}
is an Associate Professor at Northeastern University, Boston. She received her Ph.D. degree in Computer Science from the College of William and Mary, VA. She received her M.S. in Computer Science from the University of Texas at Dallas, TX and her B.S. in Computer Science from Nanjing University, China. Her research interests are capacity planning, MapReduce scheduling, cloud computing, resource management, performance evaluation, workload characterization, simulation and virtualization.
\end{IEEEbiography}

\section{Appendix1: Sequential Ratio Estimator}
\label{SEC:PROOF}

We set the bin size as $1MB$ as the prefetching size between SSD and HDD, but the original I/O request size above SSD can be any size. 
Different sizes and address distributions will impact on the workload sequential ratio, as well as the write amplification. Therefore, we need to have a module called ``stream sequentiality detector'' to estimate the sequential ratio of each disk. 
Specifically, there are two criteria in the ``sequentiality qualifying exam'':
\label{APP:SEQ}

\begin{figure}[h]
	\centering
	\includegraphics[width=0.45\textwidth]{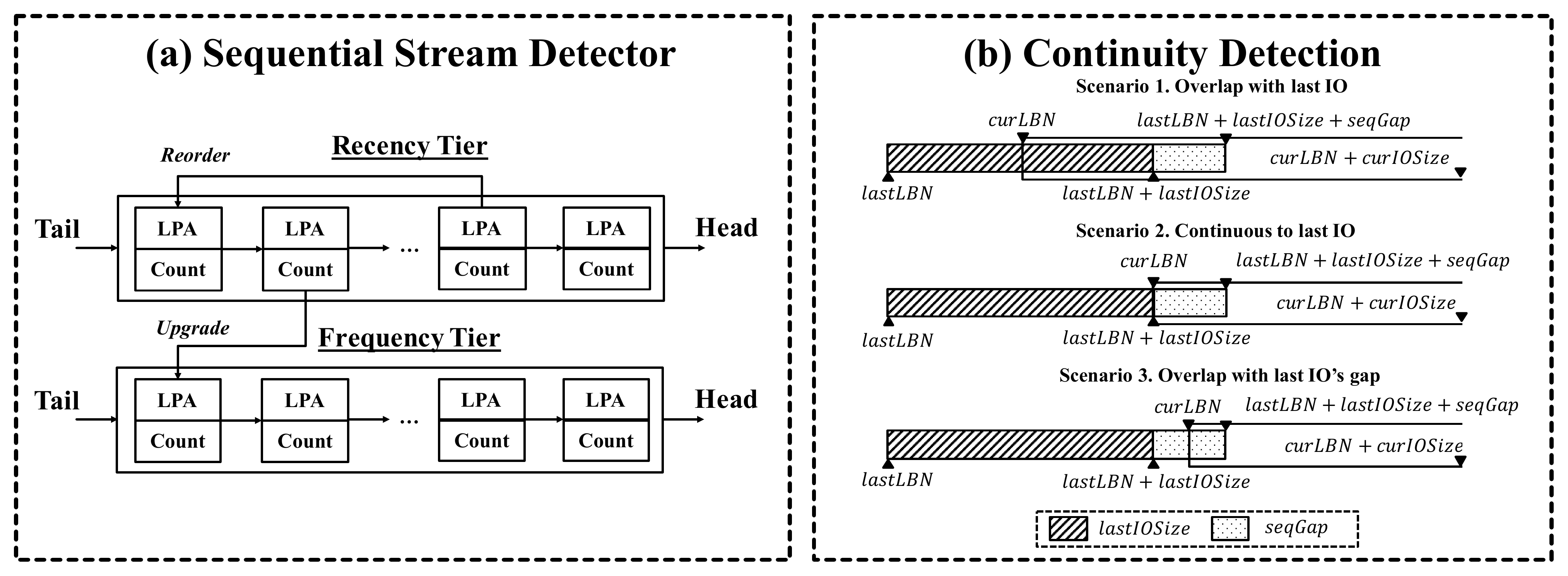}
	\caption{\small (a) Sequential stream detector and (b) Continuity detection.}
	\label{FIG:WAF_SEQ}
\end{figure}

\vspace{0.05in}
\noindent {\bf [Criterion 1] Continuity:} As shown in  Fig.~\ref{FIG:WAF_SEQ}(a), for each incoming I/O, we check whether its ``precursor neighbor'' I/Os is recorded in the queue (we have 32 queues). 
If it exists, then we update that stream node by appending this new page's metadata to its precursor neighbor node.
Otherwise, we try to create a new node in the queue for this I/O. 
The new stream I/O's node will be moved to the MRU position. 
The challenging part is to check ``whether the new I/O's precursor neighbor exists'' and ``whether the new I/O is sequential to it''. 
As illustrated in Fig.~\ref{FIG:WAF_SEQ}(b), these three scenarios are only considered that the new I/O is (sequential) successor of a certain collected stream. 
In detail, scenario 1 of Fig.~\ref{FIG:WAF_SEQ}(b) is that the current I/O's start address is within $[lastLBN, lastLBN+lastIOSize)$; 
Scenario 2 of Fig.~\ref{FIG:WAF_SEQ}(b) is a ``perfect case''  where  current I/O starts exactly from last I/O's ending point. 
In practical, due to different write granularities along the OS-to-device-firmware I/O path, it is necessary to relax the above two scenarios' conditions a little bit. 
Thus, in scenario 3 of Fig.~\ref{FIG:WAF_SEQ}(b), we provide one more ``reconsideration'' chance by extending the ending point of the last I/O by an additional space gap called ``$seqGap$'' (preset as 32 4KB-pages, i.e., 128KB).
To sum up, the continuity is determined by the relationship between the current I/O's start address ($currLBN$) and the last I/O's coverage ($lastLBN+lastIOSize$).

\vspace{0.05in}
\noindent {\bf [Criterion 2] Stream I/O Size:} Not all streams collected in the queue of  Fig.~\ref{FIG:WAF_SEQ} are considered as sequential, since I/O size is also an important criterion.
Instead of focusing on every single I/O's size, we compare each candidate stream's total I/O working set size on record (i.e., deduplicated space coverage size) with a threshold ``$seqStreamSize$'' (preset as 256 4KB-pages, i.e., $1MB$), to qualify each stream is sequential or not. 
Only streams whose total I/O working set sizes are greater than this threshold will be considered as sequential streams.

\section{Appendix2: Proof of offline mode minTCO}
\label{SEC:PROOF}

In this section, we first formulate the problem by conducting mathematical analysis on different allocation approaches, and then prove the correctness of the design to switch between grouping and greedy approaches. 

\vspace{1mm}
\noindent \textbf{\textit{Problem Formulation}}
\vspace{1mm}

We focus on the offline scenario where all disks be \textit{identical} and \textit{homogeneous}, i.e., $\mathbb{D}= \{x_i|x_i=d, i \in [1,|\mathbb{D}|]\}$, where $d$ represents a single disk, and $\mathbb{D}$ stands for the entire disk pool.
At beginning we \textit{have} the knowledge of all workloads to be deployed (i.e., the workload set $\mathbb{J}$).
We denote the total sequential ratio and the total logical data write rate of $\mathbb{J}$ as $\lambda_{\mathbb{J}}$ and $S_\mathbb{J}$, respectively.
Additionally, in real enterprise datacenters use case, the granularity of each workload is very tiny compared with that of the entire $\mathbb{J}$.
Based on these assumptions and by using the technique of mathematical induction, we start our investigate on how different approaches work and which one is better under different circumstances. 

\begin{figure}[h]
	\centering
	\includegraphics[width=0.4\textwidth]{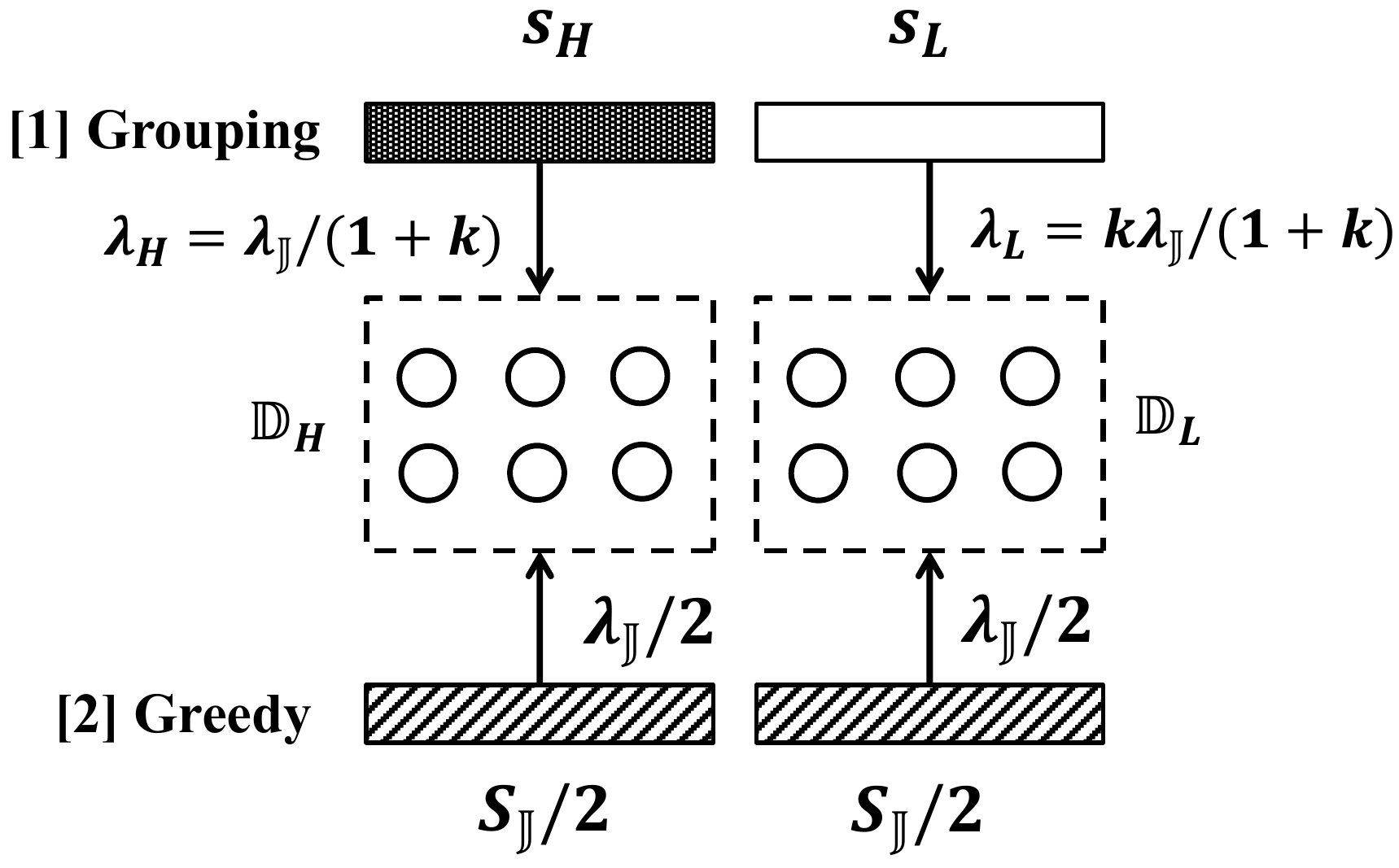} 
	\caption{\small  Grouping and Greedy approaches.}\label{FIG:MIXSEP}
\end{figure}

\noindent \textbf{\textit{Step 1: Base Case}}
\vspace{1mm}

{\mff} is an algorithm switching between two approaches, namely ``grouping'' and ``greedy'' approaches.
Fig.~\ref{FIG:MIXSEP} illustrates the example of these two approaches. Assume there are two zones with same number of disks (i.e., $\mathbb{D}_L$ and $\mathbb{D}_H$), and similar to the RAID mode estimation, we hereby regard multiple disk as one ``pseudo disk''. Let $C'_M$ and $C_I$  be the maintenance and initial costs of {\em {each zone}}, $W$  be the total cycle of {\em {each zone}}, and $A$  be the WAF function of {\em {each disk}}. 


\textbf{(1) Grouping Approach:} Workloads are sorted into two groups based on their sequential ratios (i.e., one group $\mathbb{J}_H$ with higher total sequential ratio $S_H$, and the other group $\mathbb{J}_L$ with lower total sequential ratio  $S_L$). 
Let $\lambda_H=\lambda_{\mathbb{J}}/(1+k)$ and $\lambda_L=k\lambda_{\mathbb{J}}/(1+k)$ to be the total logical write rate of these two groups, respectively, where $\lambda_H+\lambda_L=\lambda_{\mathbb{J}}$. 
According to ~\ref{SUBSUBSEC:TCO_CTM_SRW}, we have:
\small
\begin{eqnarray}
S_\mathbb{J}=\frac{\lambda_H S_H +\lambda_L S_L}{\lambda_H+\lambda_L}=\frac{1}{k+1} S_H + \frac{k}{k+1} S_L.
\label{EQ:OFFLINE-SEQ}
\end{eqnarray}
\normalsize
It then sends these two groups of workloads to different disk zones (i.e., $D_H$ and $D_L$), in order to reduce the cross-workload affect and the corresponding WAF values.
Notice that since each I/O workload is so fine-grained, we can assume that there are no capacity and throughput constrain issues during this allocation.

\textbf{(2) Greedy Approach:} It focuses on greedily filling the workload into existing disks to best use disk resources, and balancing the logical write rate of each disk, without consideration of the sequential ratio and WAF. 
It is possible to set capacity or I/O throughput usages as the balance target, however according to both the following rationale and observations from our real implementation, logical write rate plays a more important role in terms of TCO rate.
To be consistent with our analysis of the grouping approach, we also treat the greedy approach as it is sending two mixed workloads with same workload groups. Each workload group has same logical write rate $\lambda_{M}=\lambda_{\mathbb{J}}/2$, and the same sequential ratio as the entire workload set's $S_\mathbb{J}$.

\textbf{(3) TCO Comparison:} Based on the above analysis, we compare data-avg TCO rate of two approaches by using Eq.~\ref{EQ:TCO}:%
\vspace{-1mm}
\small
\begin{eqnarray}
\small &Diff\_TCO'=TCO'(Greedy)-TCO'(Grouping) \nonumber\\
\small &=\frac { 2[C_{ I }+C'_{ M } \frac { W }{ \lambda_M A(S_\mathbb{J}) } ] }{ 2[\lambda _{ M } \frac { W }{ \lambda _{ M }A( S_\mathbb{J} ) } ] }-\frac { C_{ I }+C'_{ M } \frac { W }{ { \lambda _{ H } }A(S_{ H }) } +C_{ I }+C'_{ M } \frac { W }{ \lambda _{ L }A(S_{ L }) }  }{ \lambda _{ H } \frac { W }{ \lambda _{ H }A(S_{ H }) } +\lambda _{ L } \frac { W }{ \lambda _{ L }A(S_{ L }) }  }   \nonumber\\
\label{EQ:TCODIFF2}
\small &=\frac { 2C_{ I } }{ W }  [\frac { A( S_\mathbb{J} ) }{ 2 } -\frac { 1 }{ { \frac { 1 }{ A(S_{ H }) }+\frac { 1 }{ A(S_{ L }) }  } } ]  +C'_{ M } \frac { (k-1) [A(S_{ H })-k A(S_{ L })] }{ { \lambda k [A(S_{ H })+A(S_{ L })] } }.
\label{EQ:TCODIFF}
\end{eqnarray}
\normalsize
We found that only when $k \rightarrow  1$ (i.e., for the grouping approach, its two workload groups have similar logical write rate $\lambda_H\approx\lambda_L\approx\frac{\lambda}{2}$), the second part of Eq.~\ref{EQ:TCODIFF} will be zero. Thus we have:
\small
\begin{eqnarray}
lhs\approx \frac { 2C_{ I } }{ W } \cdot [\frac { A(S_\mathbb{J}) }{ 2 } -\frac { 1 }{ { \frac { 1 }{ A(S_{ H }) } +\frac { 1 }{ A(S_{ L }) }  } } ].
\label{EQ:TCODIFF5}
\end{eqnarray}
\normalsize
Meanwhile, according to Eq.~\ref{EQ:OFFLINE-SEQ}, ``$k \rightarrow  1$'' also causes $A(S_\mathbb{J})=A(\frac { S_{ H }+S_{ L } }{ 2 } )$.  Moreover, based on our measurement and regression results (will be discussed in Sec.~\ref{SUBSUBSEC:EV_WAMM_WRM}), 
$A$ is a concave function of sequential ratio and $A\geq1$, which has the following feature:
\small
\begin{eqnarray}
A(S_\mathbb{J})=A(\frac { S_{ H }+S_{ L } }{ 2 } )\geq \frac { A(S_{ H })+A(S_{ L }) }{ 2 } .
\label{EQ:TCO_CONCAVE}
\end{eqnarray}
\normalsize
\vspace{-3mm}
\begin{figure}[h]
	\centering
	\includegraphics[width=0.4\textwidth]{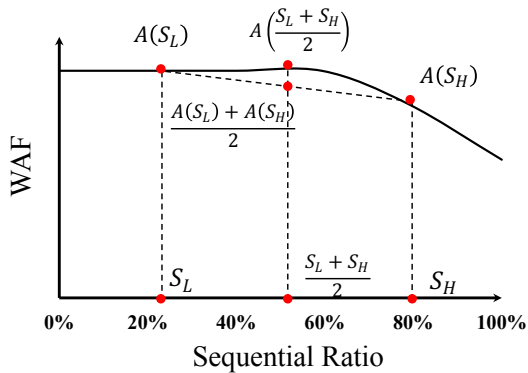}
	\caption{\small Example of NVMe WAF function.}	\label{FIG:TCO_PROOF1}
\end{figure}
\vspace{-3mm}
An example of this feature is illustrated in Fig.~\ref{FIG:TCO_PROOF1}. Furthermore, it is always true that harmonic mean is less than arithmetic mean, so we have:
\small
\begin{eqnarray}
\frac { A(S_{ H })+A(S_{ L }) }{ 2 } \geq \frac { 2 }{ \frac { 1 }{ A(S_{ H }) } +\frac { 1 }{ A(S_{ L}) }  } \geq 0.
\label{EQ:TCO_MEAN}
\end{eqnarray}
\normalsize
Consequently, by combining Eq.~\ref{EQ:TCO_CONCAVE} and ~\ref{EQ:TCO_MEAN}, we get:
\small
\begin{eqnarray}
\frac{A( \frac{S_H+S_L}{2}   )}{2}\geq \frac {1}{ \frac { 1 }{ A(S_{ H }) } +\frac { 1 }{ A(S_{ L }) }  } \geq 0.
\label{EQ:TCO_COMB}
\end{eqnarray}
\normalsize
By substituting this result into Eq.~\ref{EQ:TCODIFF}, we have $lhs\nonumber \geq 0$, only when all workloads are the same, the equal mark holds. This proves that separating sequential and random streams to different disk zones can reduce the overall TCO. 
Notice that if $k \gg 1$, we {\em cannot} guarantee that $TCO'(Greedy)$ is always worse than $TCO'(Grouping)$. 


\noindent \textbf{\textit{Step 2: Inductive Hypothesis}} 

Step 1 proves that under the certain condition ($k \rightarrow 1$), it is better to group and send workloads to two homogeneous disk zones than to mix and send them to the disk pool.
If we further divide each disk zone into two same size ``subzones'', and similarly  group each workload subset into two similar-logical-data-write-rate ($\approx\frac{\lambda_{\mathbb{J}}}{4}$) sub-subsets with relatively high and low sequential ratio workloads. It is still true that the grouping approach is better than the greedy approach.

\begin{figure}[h]
	\centering
	\includegraphics[width=0.4\textwidth]{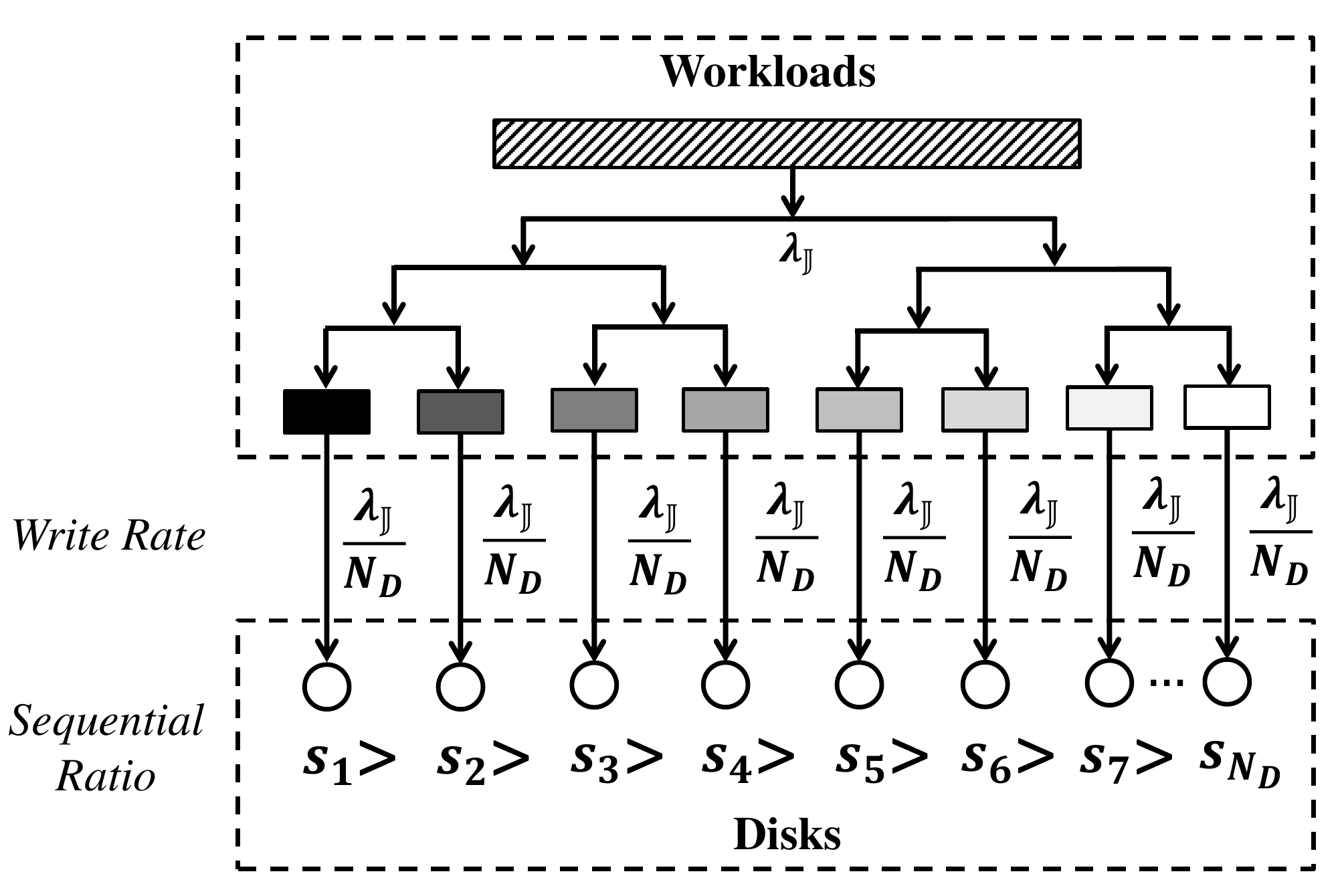}	
	\caption{\small Example of an ultimate allocation distribution of the best solution.} \label{FIG:TCO_PROOF2}
\end{figure}
\vspace{-4.5mm}

By iterating this process, we can prove that sorting all workloads and sending them to each disk by the order of different sequential ratios to balance each disk with similar write rate will lead to the best solution for minimizing data-avg TCO rate, if the workload set and homogeneous disk pool size are relatively large and workloads are fine-grained.
Fig.~\ref{FIG:TCO_PROOF2} also illustrates an example of the ultimate allocation distribution of the best solution. 

\vspace{1mm}
\noindent \textbf{\textit{Step 3: Summarize}}
\vspace{1mm}

Based on steps 1 and 2, we prove that sorting all workloads and sending them to each disk by order is the best solution. However, in real implementation, to avoid the case that purely based on sequential ratio may lead to capacity or I/O throughput unbalance, we manually set up two (or more than two) disk zones and then balance the write rate inside each zone.

\end{document}